\documentclass[aps,nofootinbib,showpacs]{revtex4}
\usepackage{graphicx}

\newcommand {\nn} {\nonumber}

\newcommand {\p} {\prime}

\newcommand{\T}{{\cal T}}
\newcommand{\ha}{\hat{\alpha}}

\newcommand{\be}{\begin{equation}}
\newcommand{\ee}{\end{equation}}
\newcommand{\bea}{\begin{eqnarray}}
\newcommand{\eea}{\end{eqnarray}}

\newcommand{\JHEP}{J. High Energy Physics}

\begin{document}

\preprint{hep-th/0202140}

\title{
Anti-de Sitter Black Holes, Thermal Phase Transition and 
Holography in Higher Curvature Gravity}

\author{Y.M. Cho ${}^{1}$ and Ishwaree P. Neupane ${}^{1,2}$}

\affiliation{${}^{1}$ School of Physics and Center for Theoretical Physics, 
Seoul National University, 151-747, Seoul, Korea\\ 
${}^{2}$ The Abdus Salam ICTP, Strada Costiera, 11-34014, Trieste, Italy\\
{\scriptsize \bf ymcho@yongmin.snu.ac.kr,  ishwaree@ictp.trieste.it }}

%\date{\today}

%% REVTEX4
%\maketitle

\begin{abstract} 
We study anti-de Sitter black holes in the Einstein-Gauss-Bonnet and the 
generic $R^2$ gravity theories, evaluate different thermodynamic 
quantities, and also examine the possibilities of Hawking-Page type thermal 
phase transitions in these theories. In the 
Einstein theory, with a possible cosmological term, one observes a 
Hawking-Page phase transition only if the event horizon is a 
hypersurface of positive constant curvature ($k=1$). But, with the 
Gauss-Bonnet or/and the $(\mbox{Riemann})^2$ interaction terms, there may 
occur a similar phase transition for a horizon of negative 
constant curvature ($k=-1$). We examine the 
finite coupling effects, and find that $N>5$ could trigger 
a Hawking-Page phase transition in the latter theory. For the Gauss-Bonnet 
black holes, one relates the entropy of the black hole to a variation 
of the geometric property of the horizon 
based on first law and Noether charge. With $(\mbox{Riemann})^2$ 
term, however, we can do this only approximately, and the two results 
agree when, $r_H>>L$, the size of the horizon is much bigger than the 
AdS curvature scale. We establish some relations between bulk data 
associated with the AdS black hole and boundary data defined on the 
horizon of the AdS geometry. Following a heuristic approach, we 
estimate the difference between Hubble entropy (${\cal S}_H$) 
and Bekenstein-Hawking entropy (${\cal S}_{BH}$) with $(\mbox{Riemann})^2$ 
term, which, for $k=0$ and $k=-1$, would imply 
${\cal S}_{BH}\leq {\cal S}_H$.   
\end{abstract}

%% REVTEX4 
\vspace*{0.3cm}

%\begin{flushleft}
%{\bf Keywords}: $R^2$ gravity, Black hole
%thermodynamics, Entropy formula \\
%\vspace*{0.2cm}


\maketitle 

\section{Introduction}
%%%%%%%%%%%%%%%%%%%%%%%%%%%%%%%%%%%%%%%%%%%%%%%%%%%%%%%%%%%%%%%%
Anti-de Sitter black hole thermodynamics, which produces an aggregate of
ideas from thermodynamics, quantum field theory and general relativity, is
certainly one of the most remarkable tools to study quantum
gravity in space-time containing a horizon~\cite{Birrell82a}. In recent
years a great deal of attention has been focused on such black holes. 
There are various familiar reasons for that. One of them, of course, is 
its role in the AdS/CFT duality~\cite{Maldacena97a}, and in particular, 
Witten's interpretation~\cite{Witten98,Witten98a} of the Hawking-Page phase
transition between thermal AdS and AdS black hole~\cite{Hawking83a} as
the confinement-deconfinement phases of the Yang-Mills (dual gauge) theory
defined on the asymptotic boundaries of the AdS geometry. It is 
possible that our observable universe is a ``brane'' living 
on the boundary of a higher-dimensional black hole, and physics of such 
black hole is holographically related to that of early (brane) universe 
cosmology~\cite{Verlinde00a}. Most of the results in the literature 
are formulated in 
terms of the AdS/CFT conjecture and based on the Einstein's theory with 
a negative cosmological constant, where one has the well known 
Bekenstein-Hawking area-entropy law
\be
\label{hawk1} {\cal S}=\frac{k_B\,c^3}{\hbar}\,\frac{A}{4G}\,,
\ee 
where $A$ is volume of the horizon corresponding to the surface at $r=r_H$.
In the following, we adopt the standard convention of
setting $c=\hbar=k_B=1$. One of the impressive features
of~(\ref{hawk1}) is its universality to all kinds of black holes~\cite
{Hawking77a,Bekenstein73a} irrespectively of their charges, shapes
and rotation. Nonetheless, it has been
known that~(\ref{hawk1}) no longer applies to the higher curvature (HC)
theories in $D>4$ dimensions~(see Ref.~\cite{Visser93a} for review).
Thus, the study of AdS black holes and boundary CFTs is well motivated 
both from the field theoretic and the cosmological points of view. 

One notes that, on general grounds, any effective gravity action will 
involve, besides the usual Einstein term, 
the higher curvatures and also derivative terms corresponding to the low 
energy matter fields(see for example~\cite{Weinberg95a}). In particular, 
when the effect of 
gravitational fluctuations are small compared to the large
number of matter fluctuations, one can neglect graviton loops, and look
for a stationary point of the combined gravitational action, and the
effective action for the matter fields. This is implied by solving
\be \label{Einstein}
R_{ab}-\frac{1}{2}\,R g_{ab}=8\pi G\, \langle T_{ab}\rangle\,,
\ee
where the source being the expectation value of the matter energy
momentum tensor, which may include the contribution from HC terms. 
If one allows non-conformally invariant matter fields, one must take 
into account the non-conformally invariant local
terms~\cite{Hawking01a}, which in four dimensions read~\cite{Duff74a,Liu98a}
\be \label{conformalT} 
\langle T\rangle =\alpha_1\, {\cal
F}-\alpha_2\,{\cal G}+\alpha_3\,\nabla^2 R\,, \ee where the
Gauss-Bonnet invariant ${\cal G}$ and the square of Weyl tensor
are \bea
{\cal G}&=&R^2-4R_{ij}R^{ij}+R_{ijkl}
R^{ijkl}\,,\nn \\
{\cal F}&=&\frac{1}{3}\,R^2-2\,R_{ij}R^{ij} +R_{ijkl}
R^{ijkl}\,.
\eea 
Here $\alpha_1,~\alpha_2,~\alpha_3$ are defined by certain combinations 
of the number of real scalars, Dirac fermions and vectors in the theory 
being considered, $\nabla^2 R$ is a variation of the local term 
$\int d^4x\sqrt{|g|}\,R^2$ in the effective action, which generally 
does not carry dynamical information. The authors 
in~\cite{Blau99a,Nojiri99a} have derived~(\ref{conformalT}) in $d+1=5$ 
for conformal anomaly in the $R^2$ gravity using the AdS/CFT 
correspondence. This suggests that higher 
curvature terms in the bulk theory can arise as next-to-leading order 
corrections in the $1/N$ (large $N$) expansion of the boundary CFTs in 
the strong 't Hooft coupling limit~\cite{Maldacena97a,Liu98a,Blau99a}. 

Paraphasing, any effective stringy gravity action also contains higher 
curvature terms of different order as loop corrections to string 
amplitudes. The most suggestive combination of higher derivative terms 
is perhaps the GB invariant, which is attributed to 
the heterotic string effective action~\cite{Deser85a,Metsaev87a}. 
The theory with a Gauss-Bonnet term is free of ghost 
when expanding about the flat space~\cite{Deser85a}. 
This is true~\cite{IPN01b} also 
in the recently discovered warped brane-world model~\cite{Randall2} 
(see~\cite{IPN00} for discussion with GB term). It is noteworthy 
that the higher derivative correction terms with ``small'' coefficients 
will not just only produce some modifications of the solution of the 
unperturbed (Einstein) theory~\cite{Simon90a}, but importantly also 
contain whole classes of new solutions in an anti-de Sitter 
space~\cite{IPN01d}, see Ref.~\cite{Germani02a} for brane-world solutions 
with the time varying gravitational constant and $\Lambda$. Moreover, 
the Einstein-Gauss-Bonnet theory in $D\geq 5$ clearly exhibits some 
new black hole solutions~\cite{Cai01a}, which are unavailable to the 
Einstein theory. 

AdS/CFT correspondence asserts that physics in the bulk of AdS 
spacetime is fully described by a CFT on the boundary, 
an intuitive notion of holography~\cite{tHooft93a}. As a result, in the 
Einstein theory with a bulk cosmological term, the thermodynamic 
quantities of the holographic dual CFTs defined on 
$S^3\times S^1$ at high temperatures may be identified with those of 
the bulk AdS Schwarzschild black holes~\cite{Witten98a}. Within the 
context of the AdS/CFT conjecture, the higher curvature terms in the bulk 
(string) theory correspond to finite coupling large $N$ effects in 
the gauge theory, one may therefore allow HC terms, which are 
squares in Ricci scalar, Ricci tensor and Riemann tensor, into the 
effective bulk action. A ghost free combination of these terms is the 
Gauss-Bonnet invariant. The higher derivative supergravity action 
including a $(\mbox{Riemann})^2$ term 
may induce some extra degrees of freedom of ghost behavior around 
stable fixed points. It is not known what this leads to the finite 
temperature field theory. Any ghost field, if exists, is 
expected to decouple in the large N limit, so they do 
not project out on the conformal boundary. However, in the full string 
theory there is no such problem. We can therefore extract a bulk 
$(\mbox{Riemann})^2$ term directly from AdS/CFT, which is necessary 
if we wish to study the finite coupling effects. 

If the AdS black hole horizon is a hypersurface with zero ($k=0$) or 
negative ($k=-1$) curvature, the black hole is always stable and 
the corresponding boundary field theory at finite temperature is 
dominated by the black hole. While, for the horizon of 
a positive curvature ($k=1$), one sees a Hawking-Page phase 
transition~\cite{Witten98a}, see also~\cite{Mueck01a}. 
This is indeed an observation one can make in the Einstein's theory, 
where the qualitative features of Hawking-Page phase 
transitions are independent of the dimensions. However, in 
the higher derivative theory with a Gauss-Bonnet term or 
a $(\mbox{Riemann})^2$ term, the situation could be different, where 
phase structures actually depend on the number of spatial dimensions 
$d$ and the horizon geometry $k$. As we have no good reasons that a dual 
description should exist for the Gauss-Bonnet theory, it would be of more 
interest to study the AdS black hole solutions with a 
$(\mbox{Riemann})^2$ term and to examine the possibility of the 
Hawking-Page phase transition. 
Wald has shown~\cite{Wald93a} that one can relate the variations 
in properties of the black hole as measured at horizon to the 
variations of the geometric property of
the horizon based on the first law and evaluation of the Noether
charge~(see Refs.~\cite{Iyer94a,Jacobson95a,Myers98a} for a clear
generalization). This has been realized for the Gauss-Bonnet black hole,
in that case we also have the exact solutions. But with a 
$(\mbox{Riemann})^2$ term, this prediction is only a good
approximation. In particular, the two entropies one finds with the 
$(\mbox{Riemann})^2$ term closely agree in the 
limit $r_H>>L$, while they completely agree for $k=0$.

The paper is organized as follows. In next section we shall
begin with our effective action and present some curvature 
quantities for a general metric ansatz. Section III deals 
in detail with the Gauss-Bonnet black hole thermodynamics in 
AdS space, including thermal phase structures. In section IV we 
study AdS black holes with a trivial $(\mbox{Riemann})^2$ 
interaction ($\gamma=0$). 
In section V we begin our discussion of the black hole 
thermodynamics with a non-zero $\gamma$, and present 
formulas for free energy, entropy and energy. We also briefly 
discuss about the finite coupling effects. In section VI, we 
present certain realizations of the FRW-type brane equations, and 
also make a comparison between the Bekenstein-Hawking entropy and 
the Hubble (or holographic) entropy. Section VII contains 
conclusions.

\section{Action, metric ansatz and curvature quantities}

Perhaps, a natural tool to explore the AdS/CFT is to 
implement the general higher 
derivative terms to the effective action, 
and to study thermodynamics of the anti-de Sitter black holes. 
To begin with, we
consider the following $(d+1)$ dimensional gravitational action, 
containing terms up to quadratic in the curvatures, 
\be\label{action1} 
I=\int
d^{d+1}x\,\sqrt{-g_{d+1}}\,\left[\frac{R}{\kappa_{d+1}}-2\Lambda
+\alpha\,R^2+\beta\,R_{ab}R^{ab}+\gamma\,R_{abcd}R^{abcd}+\cdots\right]
+ \frac{2}{\tilde{\kappa}} \int_{\partial{\cal B}} d^dx
\sqrt{|g_{(d)}|}\,{\cal K}+\cdots\,, \ee where
$\kappa_{d+1}=16\pi G_{d+1}$, ${\cal K}={\cal K}_a^a$ is the
trace of the extrinsic curvature of the boundary, ${\cal
K}^{ab}=\nabla^a n^b$, where $n^a$ is the unit normal vector on
the boundary. The last term above is attributed to
the Gibbon-Hawking boundary action. When working in
$(d+1)$-dimensional anti-de Sitter space $(\Lambda<0)$, one may drop the
surface terms including the Gibbon-Hawking action. However, the surface 
term, including the higher order, might be essential to 
evaluate the conserved quantities when the solutions are
extended to de Sitter (dS) spaces~\cite{Klemm01a,Vijay01a}.

We define the metric ansatz in the following form
\be \label{HDbrane3}
ds^2=-e^{2\phi(r)}\, dt^2 + e^{-2\phi(r)}\,dr^2
+r^2\,\sum_{i,j}^{d-1}h_{ij}\,dx^i dx^j\,,
\ee
where $h_{ij}$ is the horizon metric for a manifold ${\cal M}^{d-1}$ with 
the volume $V_{d-1}=\int d^{d-1}x\,\sqrt{h}\,$.
For~(\ref{HDbrane3}), the non-vanishing components of the
Riemann tensor are
\bea
R_{trtr}&=&e^{2\phi(r)}\left(\phi^{\p\p}+2 {\phi'}\,^2\right)\,,
\quad R_{titj}=e^{4\phi(r)}\,r\phi'\,h_{ij}
=- e^{4\phi(r)}\,R_{rirj}\,,\nn \\
R_{ijkl}&=& r^2\,{\cal R}_{ijkl}(h)-r^2\,e^{2\phi(r)}
\left(h_{ik}h_{jl}-h_{il}h_{jk}\right)\,.
\eea
We readily obtain the following non-trivial components of the Ricci tensor
\bea
R_{tt}&=& - e^{4\phi(r)}\,R_{rr}=e^{4\phi(r)}
\left(\phi''+2{\phi'}\,^2+\frac{(d-1)\phi'}{r}\right)\nn\\
R_{ij}&=& {\cal R}_{ij}(h)- e^{2\phi(r)}
\left((d-2)+2r\phi'\right)\,h_{ij}\,,
\eea
with $k$ being the curvature 
constant, whose value determines the geometry of the horizon. The
boundary topology of the Einstein space (${\cal M}^{d-1}$) looks
like
\bea\label{HDbrane7}
k=~1~\rightarrow S^{d-1}&:& {\rm Euclidean\; de\; Sitter\; space\;
(sphere)\;}\nn\\
k=~0~\rightarrow I\!\!R^{d-1}&:& {\rm flat\; space\;}\nn\\
k=-1~\rightarrow H^{d-1}&:& {\rm anti-de\; Sitter\; space\;
(hyperbolic)\;}.
\eea
This means that event horizon of the black hole can be a hypersurface
with positive,
zero or negative curvature. For spherically symmetric black holes, 
the event horizon is generally a spherical surface with $k=1$. While, if
the horizon is
zero or negative constant hypersurface, the black holes are referred as
topological black holes. The thermodynamics of topological or 
asymptotically anti-de Sitter black holes in the Einstein gravity was 
investigated, e.g., in Refs.~\cite{Brown94a,Mann96a,Vanzo97a,Birmingham98a,
Emparan99b}. 

Before going ahead, let us assume that the $(d+1)$-dimensional spacetime
is an Einstein space
\be
R_{abcd}=-\frac{1}{\ell^2}\left(g_{ac}g_{bd}-g_{ad}g_{bc}\right)\,,
\quad R_{ab}=-\frac{d}{\ell^2}\,g_{ab}\,.
\ee
This is always possible provided that the horizon geometry is also 
an Einstein space~\cite{Birmingham98a}
\be
{\cal R}_{ijkl}(h)=k\left(h_{ik}h_{jl}
-h_{il}h_{jk}\right)\,,\quad
{\cal R}_{ij}(h)=(d-2)k\,h_{ij}\,.
\ee
Then one easily computes the Ricci scalar in $(d+1)$ spacetime dimensions
\bea \label{Rscalar}
R&=&\frac{(d-1)(d-2)k}{r^2}-e^{2\phi(r)}
\left(2{\phi}^{\p\p}+4{\phi^\p}^2+\frac{4(d-1)\phi^\p}{r}
+\frac{(d-1)(d-2)}{r^2}\right)\nn \\
&=&\frac{(d-1)(d-2)k}{r^2}-\frac{1}{r^{d-1}}\,
\left(r^{d-1}\,e^{2\phi(r)}\right)^{\p\p}\,.
\eea

\section{Gauss-Bonnet black hole in AdS space} 
Let us set at first $\alpha=-\beta/4=\gamma$ in~(\ref{action1}), and
also drop the Hawking-Gibbon term. Then the
equations of motion following from~(\ref{action1}) simply read
\bea \label{GBeqn}
&&\kappa_{d+1}^{-1} \left(R_{ab}-\frac{1}{2}\,g_{ab} R\right)+\Lambda\,
g_{ab}-\frac{\alpha}{2}\, g_{ab}\,{\cal R}_{GB}^2 \nn \\
&&+2\alpha\Big(R R_{ab}-2R_{acbd}R^{cd}
+R_{acde}R_b\,^{cde}-2R_a\,^c R_{bc}\Big)=0\,
\eea
with the Gauss-Bonnet invariant
${\cal R}_{GB}^2=R^2-4R_{ab}R^{ab}+R_{abcd}R^{abcd}$. The explicit form of
the metric solution following from~(\ref{GBeqn}) is
(see also the Refs.~\cite{Deser85a,Myers88a,Wiltshire88a,Cai01a})
\be \label{adsBH5}
e^{2\phi}= k+\frac{r^2}{2\hat{\alpha}}+\epsilon\,
\frac{r^2}{2\hat{\alpha}}\left[1-\frac{4\ha}{\ell^2}
\left(1-\frac{\hat{\alpha}}{\ell^2}\right)
+\frac{4\ha\,m}{r^d}\right]^{1/2}\,,
\ee
where $\epsilon=\mp 1$, $\hat{\alpha}=(d-2)(d-3)\alpha\,\kappa_{d+1}$
and $m$ is an integration
constant with dimensions of $(\mbox{length})^{d-2}$, which is
related to the ADM mass $M$ of the black hole via
\be
M=\frac{(d-1)\,V_{d-1}}{\kappa_{d+1}}\,m\,.
\ee
We should note that the AdS curvature squared 
$\ell^2~(\equiv -\ell_{dS}^2)$ is related to the cosmological
constant $\Lambda$ via
\be \label{cosmo.const}
\Lambda=-\frac{d(d-1)}{2\kappa_{d+1}\,l^2}\,, \quad 
\mbox{where} \quad
\frac{1}{l^2}= \frac{1}{\ell^2} \left(1-\frac{\hat{\alpha}}
{\ell^2}\right)\,,
\ee
so that $\ell^2>0$ for $\Lambda<0$ (anti-de Sitter), while
$\ell^2<0$ for $\Lambda>0$ (de Sitter). Notice that there are two
branches in the solution~(\ref{adsBH5}), because $e^{2\phi(r)}$ is
determined by solving a quadratic equation, we denote by $r_+$ the
$\epsilon=-1$ branch, and by $r_-$ the $\epsilon=+1$ branch. For large $r$
\bea
-g_{00}(r_+)&\sim& k-\frac{m}{r^{d-2}}+\frac{r^2}{l^2}
+\hat{\alpha}\,{\cal O}(r^{4-2d})\,,\label{perturb1}\\
-g_{00}(r_-)&\sim& k+\frac{m}{r^{d-2}}-\frac{r^2}{l^2}
+\frac{r^2}{\ha}+\hat{\alpha}\,{\cal O}(r^{4-2d})\,.
\label{perturb2} \eea 
The $\epsilon=-1$ branch gives a $(d+1)$-dimensional 
Schwarzschild anti-de Sitter (SAdS) solution. While, $\epsilon=+1$ branch 
gives the SAdS solution, since $l^2>\hat{\alpha}$, but with a 
negative gravitational mass~\cite{Wiltshire88a}. However, $\epsilon=+1$ 
branch is unstable for certain parameter values of $\hat{\alpha}$ and 
$m$~\cite{Deser85a}. In this paper we only consider~(\ref{adsBH5}), 
rather than its perturbative cousins~(\ref{perturb1}, \ref{perturb2}).

\subsection{Thermodynamic quantities}

From~(\ref{adsBH5}), the mass of the black hole $M$ can be expressed in terms
of the horizon radius $r_+=r_H$ 
\be \label{BHmass}
M=\frac{(d-1)\,V_{d-1}\,r_+^{d-2}}{\kappa_{d+1}}
\left[k+\frac{r_+^2}{\ell^2}\left(1-\frac{\hat{\alpha}}{\ell^2}\right)
+\frac{\hat{\alpha} k^2}{r_+^2}\right]\,.
\ee 
The positions of the horizons may be determined as the real roots of the
polynomial $q\,(r=r_H)=0$, where
\be \label{qroots}
q(r)=k\,r^{d-2}+\hat{\alpha}\,k^2\,r^{d-4}+\frac{r^d}{l^2}-m\,.
\ee
Of course, in the limit $l^2\to \infty$ (i.e. $\Lambda\to  0$) and $k=1$,
one recovers the results in~\cite{Myers88a}. One derives
$q(r)=0$ directly from Eq.~(\ref{adsBH5}) or from the field
equations by setting $e^{2\phi(r)}=0$ at the horizon $r=r_+$, but one 
satisfies $e^{2\phi(r)}>0$ for $r>r_+$. When $d+1=5$, the black hole 
horizon is at
\be \label{rplusforall}
r_+^2=\frac{l^2}{2}\left[-k+\sqrt{k^2+\frac{4(m-\hat{\alpha}\,k^2)}{l^2}}
\right]\,.
\ee
There is a mass gap at $m=\hat{\alpha}k^2$, so all black holes have a 
mass $M\geq 3V_3\hat{\alpha}k^2/\kappa_5\equiv M_0$~\cite{Cai01a}, the
requirement $M> M_0$ is needed to have a black hole interpretation. 
From Eq.~(\ref{adsBH5}) and Eq.~(\ref{qroots}),
we easily see that the roots of $q(r)$ must also satisfy
\be \label{constraint}
1+\frac{4\hat{\alpha}^2\,k^2}{r^4}
+\frac{4\hat{\alpha}\,k}{r^2}\geq 0\,.
\ee
When $k\geq 0$, this is trivially satisfied in any spaces
($\ell^2>0$ (AdS), $\ell^2<0$ (dS), and $\ell^2=\infty$ (flat)), since
$\hat{\alpha}>0$. If the horizon of the black hole is a hypersurface
with a negative curvature ($k=-1$), the constraint~(\ref{constraint}) 
implies $r^2\geq 2\hat{\alpha}$ at $r=r_+$, giving the minimum 
size of the black hole horizon. For $k=+1$, 
Eq.~(\ref{constraint}) gives $r_+^2\geq -2\hat{\alpha}$, as 
first noticed in~\cite{Myers88a}. For $k=0$, $r_+$ is not constrained 
in terms of $\hat{\alpha}$.

To study the black-hole thermodynamics, it is customary to 
find first the Euclidean action by analytic continuation.
After Wick-rotating the time variable $t\to i\tau$, one
regularizes Euclidean section ${\cal E}$ by identifying
the Killing time coordinate with a period $\tau= \beta$. In the 
standard approach~\cite{Hawking83a,Horowitz95a,Witten98a}, one subtracts 
energy of the reference geometry which is simply anti-de Sitter 
space produced by a setting $m=0$ in~(\ref{adsBH5}), i.e. $M=0$. 
This process is equivalent to subtracting off the zero-temperature 
free energy in the field theory calculation. One may add the 
counter terms of holographic normalization instead of 
subtracting the energy of a reference geometry. In the 
latter approach the number of holographic terms grow with the spacetime 
dimensions, thus for simplicity we follow~\cite{Witten98a}. 
The Euclidean action $\widehat{I}$ therefore reads 
\be
\widehat{I}=-\frac{V_{d-1}\,r_H^{d-4}}{\kappa_{d+1}\,(d-3)}
\left[(d-1)\beta \left(k\,r_H^2-\hat{\alpha}\,k^2\right)
-8\pi r_H^3+3(d-1)\beta\, \frac{r_H^4}{\ell^2}\left(1-\frac{\hat{\alpha}}
{\ell^2}\right)\right]\,,
\ee
where $\beta=1/T$ is the periodicity in Euclidean time. With
$\ell^2= \infty$ (i.e. $\Lambda=0$) and $k=1$, we correctly reproduce the
result in~\cite{Myers88a}.
Since the temperature of the black hole horizon is 
identified by the periodicity in imaginary time of the metric, $T$ defines 
Hawking temperature ($T_H$) of the black hole given by 
$\left(e^{2\phi(r)}\right)^\p\arrowvert_{r=r_+}
=4\pi\,T_H$. Hence
\be \label{HawkingT}
\frac{1}{\beta}=T_H=\frac{(d-2)}{4\pi\, r_+}\,
\frac{1}{\left(r_+^2+2\hat{\alpha}\,k\right)}
\left[ k\,r_+^2 
+\frac{d-4}{d-2}\,\hat{\alpha}\,k^2
+\frac{d}{d-2}\,\frac{r_+^4}{\ell^2}
\left(1-\frac{\hat{\alpha}}{\ell^2}\right)\right]\,.
\ee 
%%%%%%%%%%%%%%%%%%%%%%%%%%%%%%%%%%%%%%%%%%%5
\begin{figure}
%\centerline{\epsfig{file=gbblack1.eps,width=10cm}}
\includegraphics[width=10cm]{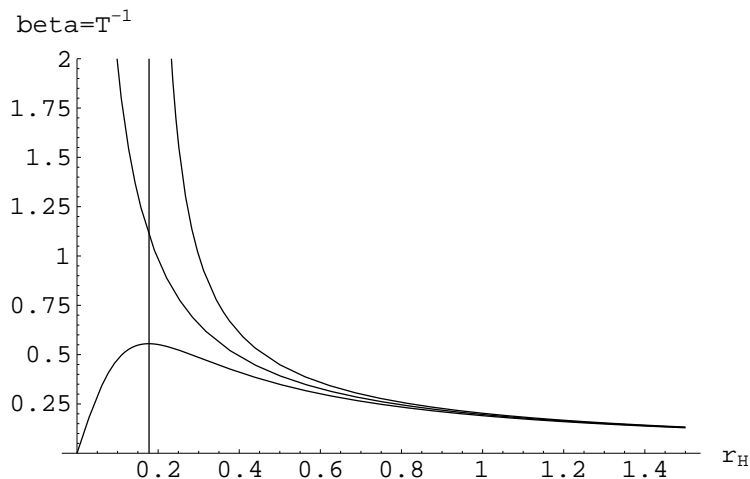}
\caption{\it Einstein gravity $(\hat{\alpha}=0$): The inverse
temperature ($\beta_0$)
versus the horizon radius ($r_H$). The three curves above from up to down
correspond respectively to the cases $k=-1$, $k=0$ and $k=+1$. The vertical 
line passing through $x\approx 0.18$ asymptotes to the $k=-1$ curve.}
\label{figure1}
\end{figure}

\begin{figure}
%\centerline{\epsfig{file=gbblack2.eps,width=10cm}}
\includegraphics[width=10cm]{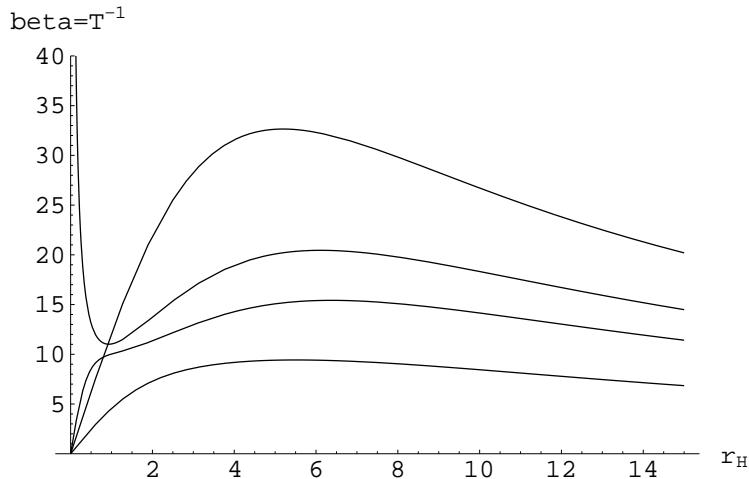}
\caption{\it Gauss-Bonnet black hole: inverse temperature
($\beta_0$) versus horizon radius ($r_H$) for $k=1$
(from up to down: $d+1=4,~5,~6$, and $10$). We have
fixed $\hat{\alpha}/\ell^2=(d-2)(d-3)\alpha\kappa_{d+1}/\ell^2$ at
$\alpha\kappa_{d+1}/\ell^2=(0.2)/81$. For $d+1=5$, a new phase of 
locally stable small black hole is seen, while 
for $d\neq 4$ the thermodynamic behavior is qualitatively similar to
that of $\hat{\alpha}=0$ case.}\label{figure2}
\end{figure}
%%%%%%%%%%%%%%%%%%%%%%%%%%%%%%%%%%%%%%%%%
We plot in Fig.~(\ref{figure1}) the inverse temperature ($\beta$) of the 
black hole versus the horizon radius $r_+$ for $\hat{\alpha}=0$ and 
$k=0,\pm1$. Then, in Fig.~(\ref{figure2}) we plot $\beta$ vs. $r_+$ 
for $\hat{\alpha}\neq 0$ and $k=1$, when the number of spatial dimensions 
$d$ is $3,\,4,\,5,$ and $9$. In term of the black hole temperature, 
we clearly see that only in $d+1=5$ and for $k=1$ there occurs a new phase 
of locally stable small black hole. In the 
Einstein gravity, one simply discards the region $\beta\to 0$
as $r\to 0$, because thermodynamically it is an unstable 
region~\cite{Hawking83a,Witten98a}. For the EGB gravity in $d=4$, 
however, both the conditions: (i) 
$\beta\to \infty$ as $r\to 0$ and (ii) $\beta\to 0$ as $r\to \infty$ 
are physical.  

We may identify the Euclidean action with the
free energy times $\beta$. Therefore 
\bea \label{GBfreeE}
F&=&-\frac{V_{d-1}\,r_+^{d-4}}{\kappa_{d+1}\,(d-3)}
\left[(d-1) \left(k\,r_+^2-\hat{\alpha}k^2\right)
-8\pi r_+^3\,T_H+3(d-1)\,\frac{r_+^4}{l^2}\right]\nn\\
&=& \frac{V_{d-1}\,r_+^{d-4}}{\kappa_{d+1}\,(d-3)}
\frac{1}{(r_+^2+2\hat{\alpha}k)}\bigg[
(d-3) r_+^4\left(k-\frac{r_+^2}{l^2}\right)-\frac{6(d-1)\,\hat{\alpha}k\,
r_+^4}{l^2} \nn \\
&{}& ~~~~~~~~~~~~~~~~~~~~~~~~~~~~~~~~~~~~~~~~~~
+(d-7)\,\hat{\alpha}k^2 r_+^2+2(d-1)\,\hat{\alpha}^2k^3\bigg] \,.
\eea 
In the second line above we have substituted the value of $T_H$
from Eq.~(\ref{HawkingT}). Interestingly, the free
energy~(\ref{GBfreeE}) was obtained in Ref.~\cite{Cai01a} by using
the thermodynamic relation $F=M-T{\cal S}$, where entropy ${\cal
S}$ was evaluated there using ${\cal S}=\int_0^{r_+} T^{-1}\,dM$.
So these two apparently different prescriptions for calculating 
free energy ($F$ read from the Euclidean action and $F$ derived 
from the first law) give the same results. One therefore computes the
energy \be E=\frac{\partial \widehat{I}}{\partial \beta_0}=M\,,
\ee where $M$ is still given by~(\ref{BHmass}), and the entropy
\be \label{entropyGB} {\cal
S}=\beta_0\,E-\widehat{I}=\frac{4\pi\,V_{d-1}\,r_+^{d-1}}{\kappa_{d+1}}
\left[1+\frac{(d-1)}{(d-3)}\,\frac{2\hat{\alpha}
k}{r_+^2}\right]\,. \ee
The minimum of the Hawking temperature is given by
solving $\partial T_H/\partial r_+=0$, where
\be\label{Tminima} \frac{\partial T_H}{\partial
r_+}=\frac{1}{4\pi}\,\frac{1}
{\left(r_+^2+2\hat{\alpha}\,k\right)^2} \left[-(d-2)\,k\,
r_+^2+\frac{d\,r_+^4}{l^2} +\frac{6d\,\hat{\alpha}k\,r_+^2}{l^2}
-(d-8)\,\hat{\alpha}\,k^2-\frac{2(d-4)\,\hat{\alpha}^2\,k^3}
{r_+^2}\right]\,. \ee
If $\hat{\alpha}=0$, for $k=0$ and $k=-1$, one easily
sees that there is no minimum of temperature, thus $k=0$ and $k=-1$ black
holes exist for all temperatures. But the situation could be different 
for $\hat{\alpha}\neq 0$. In next subsections we implement these results 
to investigate the thermal phase transition between AdS black hole and 
thermal AdS space.

\subsection{Phase Transition in the Einstein Gravity}

A black hole at high temperature is stable, while it is 
unstable at low temperature, and there can occur a phase 
transition between thermal AdS and AdS black hole at some critical 
temperature. In~\cite{Hawking83a}, 
existence of the first order phase transition was interpreted 
in terms of quantum gravity rather than boundary conformal field theory. 
Only recently, Witten~\cite{Witten98a} has interpreted this behavior as the 
confinement-deconfinement transition in dual Yang-Mills (gauge) theory. 
In Witten's interpretation the 
thermodynamics of the black hole corresponds to the thermodynamics of 
the strongly coupled
super-Yang-Mills (SYM) theory in the unconfined phase, while the
thermal anti-de Sitter space corresponds to the confined phase of the
gauge theory. The results in Einstein's theory are obtained by 
setting $\alpha=0$. Therefore 
\bea 
T_H&=&\frac{(d-2)}{4\pi\,r_+}\left(k+\frac{d}{d-2}\,\frac{r_+^2}{\ell^2}
\right)\,,
\label{TforEins}
\\
F&=&\frac{V_3\,r_+^{d-2}}{\kappa_5}\left(k-\frac{r_+^2}{\ell^2}\right)\,.
\label{FforEins} 
\eea 
%%%%%%%%%%% skip this eqn %%%%%%%%%%%%% 
%In five dimensions, the black hole horizon is at
%\be \label{rHfor0alpha} r_H^2=r_1^2=\frac{\ell^2}{2}\,
%\left(-k+\sqrt{k^2+\frac{4m}{\ell^2}}\right)\,, \quad
%m=r_H^2\,\left(k+\frac{r_H^2}{\ell^2}\right)\,. 
%\ee 
%%%%%%%%%%%%%%%%%%%%%%%%%%%%%%%%%%%%%%%%%%%%%% 
The minimum of $T_H$ occurs at $r_+=\ell\,\sqrt{k/2}$, implying that
$T_{min}=\sqrt{2k}/(\pi \ell)$. A thermal AdS phase is preferred 
for $T<T_{min}$. By definition, free energy of the thermal AdS 
space is zero, but from~(\ref{FforEins}) $F=0$ only at 
$r_+=\ell\sqrt{k}$. The critical Hawking temperature is therefore
$T_H=T_c=3\sqrt{k}/{(2\pi \ell)}$. Thus, $k=0$ and $k=-1$ cases are 
of less interest to explain the phase transition. But, for $k=1$, 
$T_c$ defines a possible first order phase transition. When $T_H>T_c$ 
a stable AdS black hole exists, but a thermal AdS phase 
is preferred for $T_c>T_H$. 
%%%%%%%%%%%%%%%%%%%%%%%%%%%%%%%%%%%%%%%%%%
\begin{figure}
%\centerline{\epsfig{file=gbblack3.eps,width=10cm}}
\includegraphics[width=10cm]{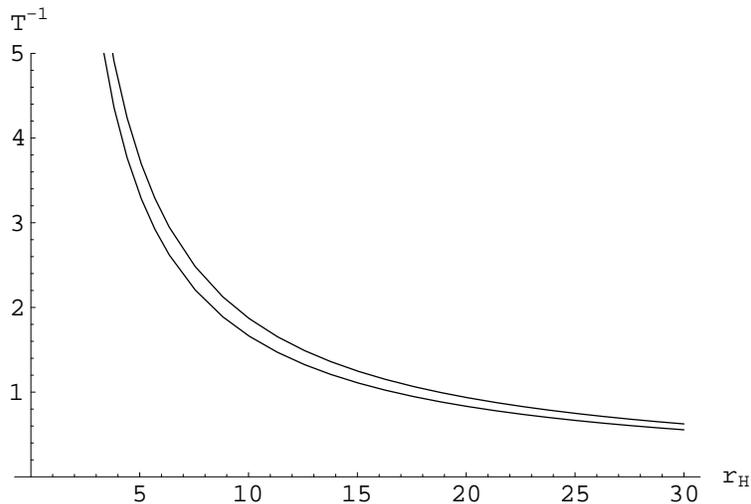}
\caption{\it Gauss-Bonnet black hole: inverse temperature vs
horizon radius for $k=0$, $d=4$, and 
$\hat{\alpha}/\ell^2=(0.7)/(0.9)^2$. Only for $\hat{\alpha}=\ell^2$, $T_H=T_c$,
and hence $F=0$, otherwise free energy is always
negative, since $\hat{\alpha}<\ell^2$ should hold for $\Lambda<0$.}
\label{figure3}
\end{figure}
%%%%%%%%%%%%%%%%%%%%%%%%%%%%%%%%%%%%%%%%%%%%%%%%%%%%%%%%%
\begin{figure}
%\centerline{\epsfig{file=gbblack4.eps,width=10cm}}
\includegraphics[width=10cm]{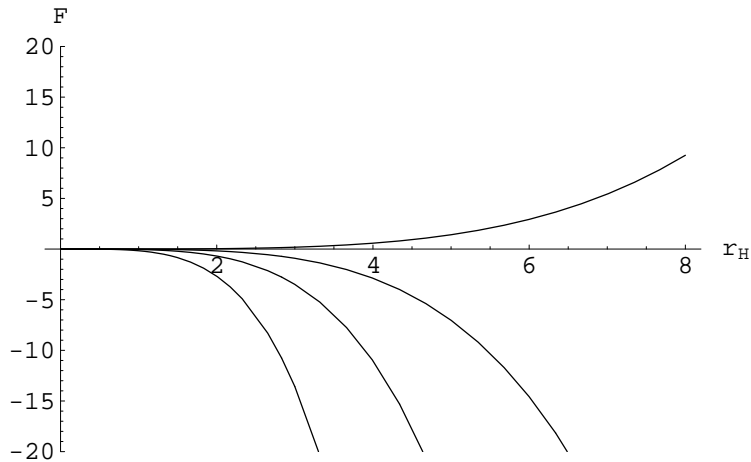}
\caption{\it Gauss-Bonnet black hole: free energy vs
horizon radius for $k=0$ and $d=4$. The curve with $F>0$ corresponds to 
$\hat{\alpha}/\ell^2=(0.7)/(0.836)^2$, which gives a dS solution, 
since $\ell^2>\hat{\alpha}$ and hence $\Lambda>0$. The other three curves 
with $F<0$ from up to down correspond respectively to 
$\hat{\alpha}/\ell^2=(0.7)/(0.84)^2,\,(0.7)/(0.85)^2$ and 
$0.7/(0.9)^2$.}
\label{figure4}
\end{figure}
%%%%%%%%%%%%%%%%%%%%%%%%%%%%%%%%%%%%%%%%%%%%%%%%%%%%%%%%
\begin{figure}
%\centerline{\epsfig{file=gbblack5.eps,width=10cm}}
\includegraphics[width=10cm]{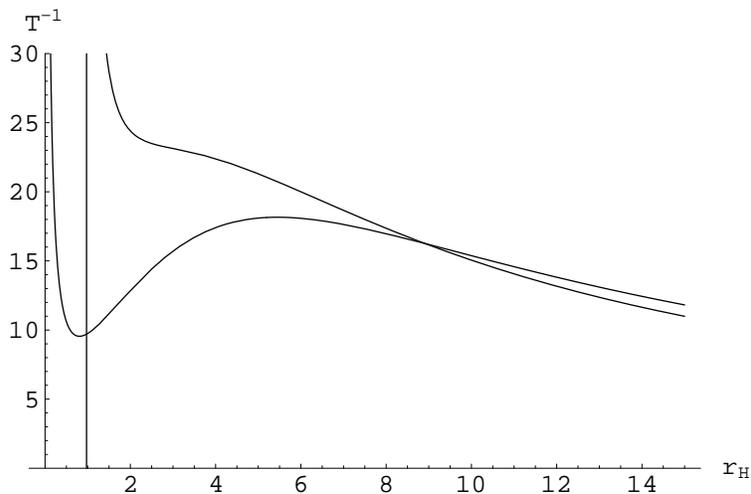}
\caption{\it Gauss-Bonnet black hole: inverse temperature vs 
horizon radius for the case $k=1$ in $d=4$ and
$\hat{\alpha}/\ell^2=(0.3)/64$. The upper curve corresponds to
$T_c^{-1}$ and the lower one to $T_H^{-1}$. The region where $T_H$ exceeds
$T_c$ is shown.}\label{figure5}
\end{figure}
%%%%%%%%%%%%%%%%%%%%%%%%%%%%%%%%%%%%%%%%%%%%%%%%%%%%%%%
\begin{figure}
%\centerline{\epsfig{file=gbblack6.eps,width=10cm}}
\includegraphics[width=10cm]{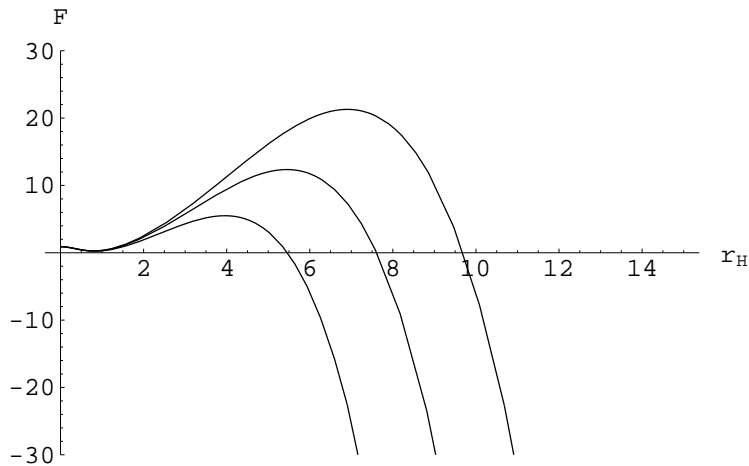}
\caption{\it Gauss-Bonnet black hole: free energy vs horizon radius 
for the case $k=1$ in $d=4$. The three curves
from up to down correspond respectively to the cases
$\hat{\alpha}/\ell^2=(0.3)/100,~(0.3)/64$ and $\hat{\alpha}/\ell^2=(0.3)/36$.
The $k=1$ is the most plausible situation for a Hawking-Page phase
transition even if $\hat{\alpha}\neq 0$.}
\label{figure6}
\end{figure}
%%%%%%%%%%%%%%%%%%%%%%%%%%%%%%%%%%%%%%%%%%%%%%%%
\begin{figure}
%\centerline{\epsfig{file=gbblack7.eps,width=10cm}}
\includegraphics[width=10cm]{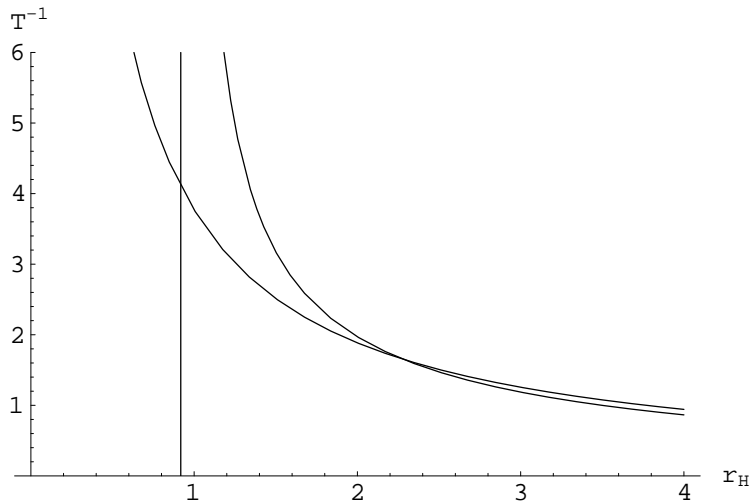}
\caption{\it Gauss-Bonnet black hole: inverse temperature vs
horizon radius for $d=4$, $k=-1$ and 
$\hat{\alpha}/\ell^2=(0.3)/(0.775)^2$, the latter choice best represents 
the average for which $F>0$, see Fig~(\ref{figure9}). 
The two curves from up to down correspond to $T_c^{-1}$ and 
$T_H^{-1}$. For a small $r$, one has $F>0$ before hitting a 
singularity at $r=r_*$ (see Fig.~(\ref{figure8})), other than at $r=0$, 
and $r_*$ is shielded by $r_H$.}
\label{figure7}
\end{figure}
%%%%%%%%%%%%%%%%%%%%%%%%%%%%%%%%%%%%%%%%%%%%%%%%%%
\begin{figure}
%\centerline{\epsfig{file=gbblack8.eps,width=10cm}}
\includegraphics[width=10cm]{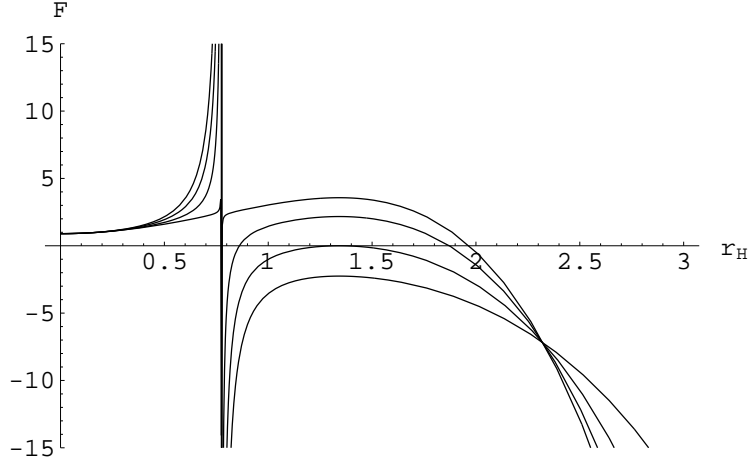}
\caption{\it Gauss-Bonnet black hole: free energy vs horizon
radius for the case $k=-1$ in $d=4$. The four curves from up to down
correspond respectively to the values
$\hat{\alpha}/\ell^2=(0.3)/(0.8)^2,~
(0.3)/(0.65)^2,~(0.3)/(0.6)^2$ and $(0.3)/4$. For $\hat{\alpha}/\ell^2>(0.3)/
(0.6)^2$ and $\hat{\alpha}/\ell^2<(0.3)/
(1.5)^2$, the free energy is always negative.}
\label{figure8}
\end{figure}
%%%%%%%%%%%%%%%%%%%%%%%%%%%%%%%%%%%%%%%%%%%%%%%%%%%%%%%
\begin{figure}
%\centerline{\epsfig{file=gbblack9.eps,width=10cm}}
\includegraphics[width=10cm]{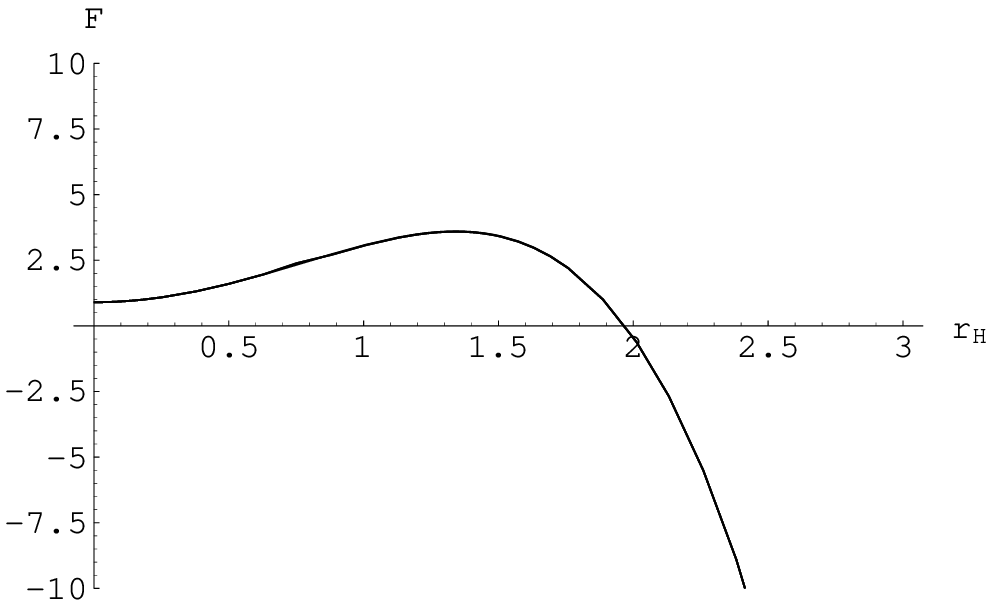}
\caption{\it Gauss-Bonnet black hole: free energy ($F$) vs
horizon radius ($r_H$) for the case $k=-1$ in $d=4$. All the curves in 
between $(0.3)/(0.7595)^2\leq \hat{\alpha}/\ell^2\leq (0.3)/(0.7905)^2$ 
coincide each other. This is the region where a small topological black 
hole might prefer the thermal AdS phase, $F>0$.}\label{figure9}
\end{figure}
%%%%%%%%%%%%%%%%%%%%%%%%%%%%%%%%%%%%%%%%%%%%%%%%%%%%%%%%%%
\subsection{Phase Transition in the Einstein-Gauss-Bonnet Gravity}
We may allow a non-trivial GB coupling 
$\alpha$ to investigate the Hawking-Page transitions in gravity side. It is
suggestive to consider the $d+1\geq 6$ case separately, since the
properties of the solutions differ from the special case
$d+1=5$, for example, see Fig.~(\ref{figure2}). We consider here 
only the $d+1=5$ case for two reasons: (i) from $AdS_5/CFT_4$ point 
of view, and (ii) other than in $d+1=5$ there is no new phase 
transition of locally stable small black hole. Then 
\be \label{freeF5}
F=-\frac{V_3}{\kappa_5}\left[3k\left(r_H^2-\hat{\alpha}\,k\right)
-8\pi r_H^3\,T_H+\frac{9
r_H^4}{\ell^2}\left(1-\frac{\hat{\alpha}}{\ell^2}
\right)\right]\,, \ee where $\hat{\alpha}=2\alpha\kappa_5$. Thus,
free energy can be zero at the critical Hawking temperature \be
\label{criticalT} T_H=T_c=\frac{1}{8\pi\,
r_H}\left[3k\left(1-\frac{\hat{\alpha}\,k}{r_H^2} \right)
+\frac{9\,r_H^2}{\ell^2}\left(1-\frac{\hat{\alpha}}{\ell^2}\right)\right]\,.
\ee The minimum of $T_H$ is given by solving 
\be
\label{rootsforT}
r_+^2\left(k-\frac{2\,r_+^2}{l^2}\right)+2\hat{\alpha}k
\left(k-\frac{6\,r_+^2}{l^2}\right)=0\,, \ee
provided $(r_+^2+2\hat{\alpha}k)\neq 0$. In the following, it would be
suggestive to consider the $k=0,\,1\,$ and $-1$ cases separately.  
\vspace*{0.2cm}

{\large\bf \underline{$k=0$ case}}: From Eq.~(\ref{HawkingT}) in 
$d=4$ and Eq.~(\ref{criticalT}), $F=0$ only if $\ell^2=\hat{\alpha}$.  
%\begin{equation} 
%1-\frac{\hat{\alpha}}{\ell^2}=0\,.
%\end{equation}
For $\ell^2>\hat{\alpha}$, since one has $F<0$, the black 
hole is always stable. For $\ell^2< \hat{\alpha}$, 
free energy~(\ref{freeF5}) appears to be positive, 
but this limit is not allowed for AdS black hole, because 
$\ell^2<\hat{\alpha}$ necessarily implies $\Lambda>0$ and the 
black hole mass in Eq.~(\ref{BHmass}) is negative. Thus 
$\ell^2\geq \hat{\alpha}$ puts a bound for flat ($k=0$) AdS
black holes. We see no evidence of a phase transition for $k=0$ even if 
$\hat{\alpha}\neq 0$. This behavior is not changed for
$(d+1)>5$, unlike the case will be with $k\neq 0$. 

{\large \bf \underline{$k=+1$ case}}: This is the most interesting 
case, since the boundary 
of the bulk manifold will have an intrinsic geometry same as the 
background of the boundary field theory at finite temperature (i.e. 
$S^1\times S^{3}$). From Eq.~(\ref{HawkingT}) in $d=4$ 
and Eq.~(\ref{criticalT}), we have $F\geq 0$, only if
\be \label{inequalT1}
\left(1-\frac{r_+^2}{l^2}\right)- \frac{3\hat{\alpha}}{r_+^2}
\left(1-\frac{2\hat{\alpha}}{r_+^2}+\frac{6\,r_+^2}{l^2}\right)\geq
0\,.
\ee
One has $F>0$ for $T<T_c$, while $F<0$ for $T>T_c$. 
As seen in Fig.~(\ref{figure6}), small Gauss-Bonnet black hole with 
a spherical horizon, which has a small positive free energy at start, 
evolves to thermal AdS phase, attains a maximum positive free energy at 
some $r=r_+$, and eventually goes to a stable black hole phase for 
large $r$. 

{\large\bf \underline{$k=-1$ case}}: In this case, the $F\leq 0$ 
condition reads 
\be
\label{inequalT2}\left(1+\frac{r_+^2}{l^2}\right)+
\frac{3\hat{\alpha}}{r_+^2}
\left(1+\frac{2\hat{\alpha}}{r_+^2}-\frac{6\,r_+^2}{l^2}\right)\geq
0\,. \ee 
%%%%%%%%%%%%%% skip these lines %%%%%%%%%%% 
%$F$ is always negative if $\hat{\alpha}=0$, thus a hyperbolic AdS 
%black hole is always stable with $\hat{\alpha}=0$. 
%However, the situation is different if $\hat{\alpha}\neq 0$. 
%%%%%%%%%%%%%%%%%%%%%%%%%%%%%%%
At the critical radius $r_+^2=2\hat{\alpha}$, free energy~(\ref{freeF5}) 
is zero when $r_+^2=\ell^2$, $F$ is always negative for
$r_+^2>2\hat{\alpha}$ and $r_+^2>\ell^2$, but $F$ can be 
positive for $2\hat{\alpha}\leq r_+^2<\ell^2$, which may trigger a 
Hawking-Page type transition even if the horizon is hyperbolic 
(see Figs.~(\ref{figure8}) and (\ref{figure9})). 
%%%%%%%%%%%% skip this equation to make short %%%%%%%%%% 
%For $k=-1$, a singularity, other than at $r=0$, at $r_*$ 
%\be
%r_*^{d}\,\left(1-\frac{4\hat{\alpha}}{l^2}\right)-4\hat{\alpha}\,r_+^{d-2}
%\left(1-\frac{r_+^2}{l^2}-\frac{\hat{\alpha}}{r_+^2}\right)=0 \,,
%\ee other than at $r=0$, is shielded by $r_+$. 
%%%%%%%%%%%%%%%%%%%%%%%%%%%%%%%%%%%%%%%%%%%%%%%%%%%%%%%% 
We note from Fig.~(\ref{figure9})
that under the limit $0.48<\hat{\alpha}/\ell^2<0.52$, free
energy is always positive. Thus $k=-1$ may allow one 
to study the boundary field theory at finite temperature with 
different background geometries~\cite{Birmingham98a,Emparan99a}. 
Due to a possible phase transition for black holes with a hyperbolic 
horizon, one may find it particularly amusing that when the hypersurface 
is $AdS_3\times S^1$ ($AdS_3$ may be obtained by analytic 
continuation of $dS_3$) quantum gravity in $AdS_5$ can be dual to 
a boundary conformal field theory on $AdS_4$ background. This 
possibly reflects that the geometry on the boundary is not dynamical, 
since there are no gravitational degrees of freedom in the dual 
CFT~\cite{Emparan99a}. 
%%%%%%%%%%%%%%%%%%%%%%%%%%%%%%%%%%%%%%%%%%%%%%%%%%%%%%%%%%%%
\section{Thermodynamic quantities with $\gamma=0$}
%%%%%%%%%%%%%%%%%%%%%%%%%%%%%%%%%%%%%%%%%%%%%%% 
For $\gamma=0$, the field
equations following from~(\ref{action1}), integrate to give the
metric solution~(\ref{HDbrane3}) with~\cite{Nojiri01a} \be
\label{HDbrane4} e^{2\phi(r)}=
k+\frac{r^2}{L^2}-\frac{\mu}{r^{d-2}}\,, \ee where 
$L^2$ is related to the cosmological term
$\Lambda$ via \be \label{HDbrane5}
\Lambda=-\frac{d(d-1)}{2\kappa_{d+1}\,L^2}\,
\left(1-\frac{(d-3)\,\varepsilon}{2(d-1)}\right)\,, \ee where \be
\label{varep1}
\varepsilon=\frac{2d\big(\beta+(d+1)\alpha\big)\,\kappa_{d+1}}{L^2}\,.
\ee The integration constant $\mu$ is related to the black
mass $M=(d-1)\,V_{d-1}\,\mu/\kappa_{d+1}$. When $\mu=0$, the
solution~(\ref{HDbrane4}) locally corresponds to $AdS$ metric, while
$\mu\neq 0$ gives AdS black hole solutions. 
For the background metric~(\ref{HDbrane4}), the classical action
takes the form \be I=-\frac{V_3}{2\pi G_5\,
T}\int_{r_H}^{\infty}dr\,r^3 \left(1-\varepsilon\right)\,, \ee
where $V_3$ is the volume of the manifold ${\cal M}^{3}$ and
$\varepsilon=8\kappa_5\left(\beta+5\alpha\right)/L^2$. After a
proper regularization, the Euclidean action is identified with
the free energy $(F)$ times $1/T_H$. 
Following~\cite{Hawking77a,Hawking83a}, the free energy $(F)$ and
the Hawking temperature $T_H$ evaluated for $d=4$ read 
\bea
F&=&\frac{V_3\,r_H^2}{16\pi G_5}\left(k-\frac{r_H^2}{L^2}\right)
\big(1-\varepsilon\big)\,,\label{freeFforg0}\\
T_H&=&\frac{1}{4\pi}\,\left[e^{2\phi(r)}
\right]^\prime\arrowvert_{r=r_H}
=\frac{1}{\pi\,L^2}\left(r_H+\frac{k\,L^2}{2r_H}\right)\,. \eea 
As in~\cite{Nojiri01a}, if one defines 
$\tilde{G}_5\,(1-\varepsilon)= G_5$, entropy and energy will have 
the usual form \be \label{energy1} {\cal
S}=-\frac{dF}{dT_H}=\frac{V_3\,r_H^3}{4\tilde{G}_5}\,, \quad
E=F+T{\cal S}= \frac{3V_3\,\mu} {16\pi\tilde{G}_5}\equiv M\,. \ee
Thus energy of the AdS black hole can be identified simply by mass
of the black hole, in terms of a renormalized Newton constant $\tilde{G}_5$.
Moreover, these quantities may be identified, up to a conformal
factor, with the same quantities defined on the boundary of $AdS_5$. 
%%% skip the following %%%%%%% 
%It is not unexpected that AdS gravity with $\gamma=0$ is 
%conformal, rather this may be related to AdS/CFT 
%trace anomaly~\cite{Blau99a}, since the leading supergravity 
%contribution to the trace anomaly involves only
%the squares of the Ricci scalar and Ricci tensor of the boundary 
%metric, not the square of the Riemann tensor.
%
\subsection{The role of boundary terms}
For a definiteness, we work in $d+1=5$, and define $e^{-\sigma(y)}=r$. Then
we can bring the metric~(\ref{HDbrane3}) in the following Randall-Sundrum
type five-dimensional warped metric~\footnote
{The related coordinate
transformations are given in Ref.~\cite{Nojiri01a}, which read:
$e^{2\phi}{\dot{t}}^2-e^{-2\sigma}e^{-2\phi}{\dot{\sigma}}^2 =
e^{-2\sigma}$; $e^{2\phi}\dot{t}\,t'-e^{-2\sigma}
e^{-2\phi}\dot{\sigma}\sigma'=0$, and
$e^{2\phi}{t'}^2-e^{-2\sigma} e^{-2\phi}{\sigma'}^2=1$, where
$\dot{\sigma}=\partial_\tau\sigma,~\sigma'=\partial_y\sigma$
\label{co-trans}.} \be \label{RSmetric}
ds^2=e^{-2\sigma(y)}\left(-d\tau^2+\sum_{i=1}^{3}\gamma_{ij}\,dx^i
dx^j\right) +dy^2\,, \ee where $y$ denotes an extra (fifth) space
transverse to the brane, which picks out a family of
hypersurfaces, $y=\mbox{const}$. To make the role of the
``brane'' dynamic, one also adds to~(\ref{action1}) the following
boundary term, corresponding to the vacuum energy on the hypersurface,
\be \label{HDbrane8} I_{\partial{\cal
B}}=\int_{\partial{\cal B}}
d^{d}x\sqrt{|g_{(d)}|}\,\left(-\T\right)+\cdots \,,
\ee where, the
ellipsoids represent the higher order surface terms, and
$\partial{\cal B}$ denotes the hypersurface with a
constant extrinsic curvature.
To the leading order, one may drop the higher
order surface terms by imposing certain restriction on their
scalar invariants (see, for example,~\cite{Nojiri01b}), and
this is the choice we adopt here. The extrinsic curvature ${{\cal
K}}_{ab}$ can be easily calculated
from~(\ref{RSmetric}). The variation $\delta I+\delta
I_{\partial{\cal B}}$ at the $4d$ boundary would rise to give \be
\label{HDbrane10} \delta I+\delta I_{\partial{\cal B}}= \int d^4x
\sqrt{|g_{(4)}|} \left[\left(\frac{8(1-\varepsilon)}
{\kappa_5}-\frac{8}{\tilde{\kappa}}\right) \,\delta \sigma' +
4\left(\frac{8}{\tilde{\kappa}}\,\sigma'
-\frac{2(1-\varepsilon)}{\kappa_5}\,\sigma' + \frac{\T}{2}\right)
\delta \sigma\right]\,. \ee Let us assume that
the gravitational couplings $\kappa$ and $\tilde{\kappa}$ are
related by~\cite{Nojiri01a} 
\be\label{twocouplings}
\frac{1}{\tilde{\kappa}}-\frac{1-\varepsilon}{\kappa_5}=0
\Rightarrow \tilde{G}_5=\frac{G_5}{(1-\varepsilon)}\,, \ee so that the
first bracket in~(\ref{HDbrane10}) vanishes. The dynamical
equations on the brane then reduce to \be \label{HDbrane11}
\T=-\frac{12(1-\varepsilon)}{\kappa_5}\,{\sigma'}|_{y=0+}
=\frac{12(1-\varepsilon)}{\kappa_5\,L}\,, \quad
\Lambda=-\frac{6}{\kappa_5
L^2}\left(1-\frac{\varepsilon}{6}\right)\,, \ee where
$\varepsilon$ is given by~(\ref{varep1}), and in the second step
$\sigma'|_{y=0+}= - 1/L$ has been used. The condition $\sigma^\p=-1/L$
seems useful to recover the RS-type fine tunings.

\section{Thermodynamic quantities with $\gamma\neq 0$}

In AdS/CFT, the higher curvature terms in the bulk have coefficients
that are uniquely determined~\cite{Liu98a,Blau99a}, thus it is suggestive to
specify the coefficient $\gamma$. In particular, one computes the trace
anomaly of a 
${\cal N}=2$ SCFT in $AdS_5$ supergravity or conformal anomaly on $S^4$ 
in the ${\cal N}=4$ super YM theory. One finds there the order $N$ 
gravitational contribution to the anomaly from a
$(\mbox{Riemann})^2$ term. The string theory dual to
${\cal N}=2$ SCFTs with the gauge group $Sp(N)$ has been
conjectured to be type IIB string theory on $AdS_5\times S^5/Z_2$ 
background~\cite{Spalinski98a,Blau99a}, whose low energy effective 
(bulk) action in five dimensions reads 
\be \label{5dAdS} 
S=\int d^5x \sqrt{-g}\left[\frac{R}{\kappa_5}-2\Lambda
+\gamma\, R_{abcd}R^{abcd}\right]+(\mbox{boundary~ terms})\,, \ee
if we define
\be \label{sugrarel.}
\frac{L^3}{\kappa_5}=\frac{N^2}{4\pi^2}\,, \quad
\Lambda=-\frac{6}{\kappa_5\,L^2}=-\frac{6\,N^2}{4\pi^2}\,\frac{1}{L^5}\,,
\quad \gamma=\frac{6N}{24\cdot 16\pi^2}\,\frac{1}{L}\,, \ee hence
$\gamma>0$. The third term in~(\ref{5dAdS}) is suppressed by $1/N$ with 
respect to the first two terms, thus taking $N\to \infty$ 
(i.e. $\alpha'=L^2/(4\pi g_{st}N)^{1/2}\to 0$, where $g_{st}$ is the 
string coupling) enables one to take classical limit of the string 
theory on $AdS_5\times S_5$~\cite{Maldacena97a}. If we wish to study finite 
$N$ coupling effects, it is necessary to take into account the leading 
higher derivative corrections, including $(\mbox{Riemann})^2$ term.    

A string background $AdS_5\times S^5$ is 
an exact solution of Type IIB theory~\cite{Maldacena97a}, where one knows 
that there is no $R^2$ term, and sub-leading corrections come only from 
$(\mbox{Weyl})^4$ term. Therefore, the motivation for including 
$(\mbox{Riemann})^2$ term in~(\ref{5dAdS}) must be clear. As worked out 
in~\cite{Blau99a}, this term arises due to the orbifold compactification to 
$AdS_5\times X_5$, where $X_5=S^5/Z_2$ for ${\cal N}=2$ gauge theory. 
Interestingly, a bulk $(\mbox{Riemann})^2$ term would rise to give 
subleading ${\cal O}(N)$-term predicted for $d=4$, ${\cal N}=2$ 
superconformal field theory with $Sp(N)$ gauge group~\cite{Blau99a}. 
We will comment upon the finite coupling effects, but before that we 
evaluate the black hole parameters for the above theory. 
%%%%%%%%%%%%%%%%%%%%%%%%%%%%%%%%%%%%%%%%%%%%%%%%%%
\subsection{Black hole solutions}
\begin{figure}
%\centerline{\epsfig{file=gbblack10.eps,width=10cm}}
\includegraphics[width=10cm]{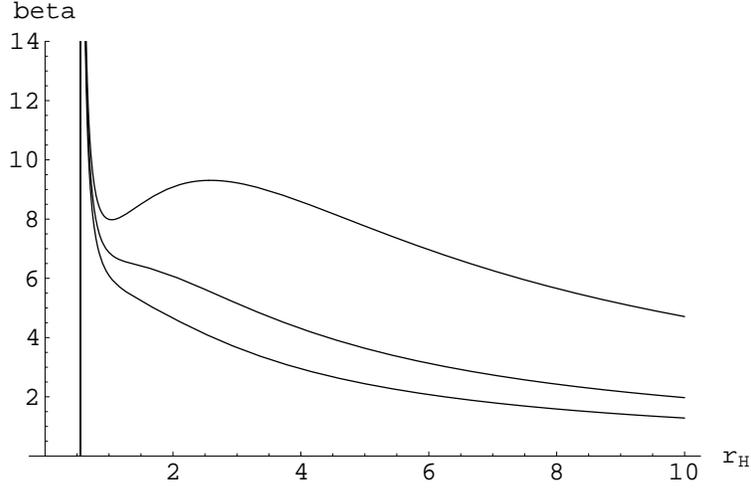}
\caption{\it Black hole in $(\mbox{Riemann})^2$ gravity: inverse 
temperature vs horizon radius ($r_H$) for $k=+1$ in $d=4$. The 
three curves from up to down correspond respectively to 
$\hat{\gamma}/L^2=(0.15)/(4.0)^2,\,(0.15)/(2.511)^2$ and $(0.15)/(2.0)^2$}
\label{figure10}
\end{figure}

\begin{figure}
%\centerline{\epsfig{file=gbblack11.eps,width=10cm}}
\includegraphics[width=10cm]{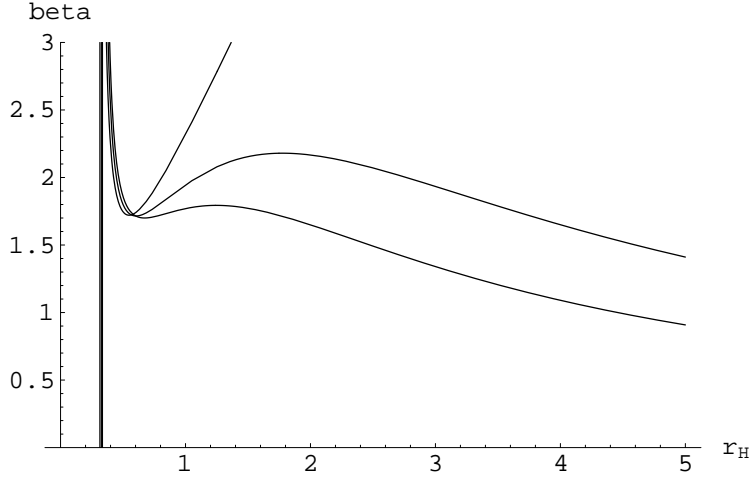}
\caption{\it Black hole in $(\mbox{Riemann})^2$ gravity:
inverse temperature vs horizon radius for $k=-1$ in $d=4$. 
The three curves from up to down correspond respectively to 
$\hat{\gamma}/L^2=(0.15)/(0.3875)^2,\,(0.15)/(0.4)^2$ and $(0.15)/(0.41)^2$.  
Black holes with $\hat{\gamma}/L^2>(0.15)/(0.3875)^2$ (i.e. 
$\hat{\gamma}>3\ell^2$ or $\Lambda>0$) are unstable. A line asymptotic 
to each curve shifts towards right as $\hat{\gamma}/L^2$ is decreased, 
hence a bold asymptote appears.} 
\label{figure11}
\end{figure}

\begin{figure}
%\centerline{\epsfig{file=gbblack12a.eps,width=10cm}}
\includegraphics[width=10cm]{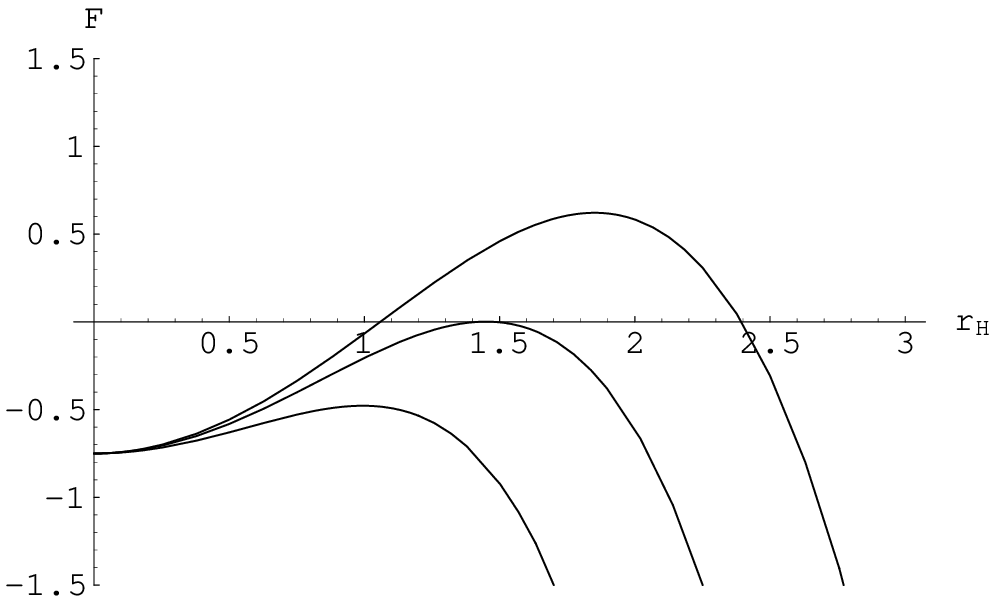}
\caption{\it Black hole in $(\mbox{Riemann})^2$ gravity:
free energy vs  horizon radius for $k=+1$ in $d=4$. 
The three curves from up to down correspond respectively to 
$\hat{\gamma}/L^2=(0.15)/(3.0)^2,~(0.15)/(2.511)^2$ and 
$(0.15)/(2.0)^2 $.}\label{figure12}
\end{figure}

%%%%%%%%%%%%%%%%%%%%%%%%%%%%%%%%%%%%%%%%% 
When $\gamma\neq 0$, we can find only a perturbative 
metric solution, which reads, in $(d+1)=5$, 
\be
\label{higher1a}
e^{2\phi(r)}= k-\frac{\mu}{r^2}
\left(1+\frac{2\hat{\gamma}}{\ell^2}\right) +\frac{r^2}{\ell^2}
\left(1+\frac{2\hat{\gamma}}{3\,\ell^2}\right)
+\frac{\hat{\gamma}\,\mu^2}{r^6} +{\cal O}(\hat{\gamma})^2\,, \ee
where $\hat{\gamma}=2\gamma\,\kappa_{5}$.
Since $\mu$ is an integration constant with dimension of
$(\mbox{length})^{d-2}$, we may rescale \be \mu
\left(1+\frac{2\hat{\gamma}}{\ell^2}\right)\to \tilde{M}\,,\quad
\frac{1}{\ell^2}\left(1+\frac{2\hat{\gamma}}{3\,\ell^2}\right)\to
\frac{1}{L^2}\,, \ee in order to bring the solution~(\ref{higher1a})
in the usual form
\be \label{higher1} 
e^{2\phi(r)}=
k-\frac{\tilde{M}}{r^2}+\frac{r^2}{L^2}
+\frac{\hat{\gamma}\,\tilde{M}^2}{r^6} +{\cal
O}\big(\hat{\gamma}^2\big)\,. \ee 
The integration constant $\tilde{M}$ at the singularity
$e^{2\phi(r)}=0$ reads, from Eq.~(\ref{higher1}), as \be
\label{higher4} \tilde{M}_+=
r_+^2\left[k+\frac{r_+^2}{L^2}+\frac{\hat{\gamma}}{r_+^2}
\left(k^2+\frac{2kr_+^2}{L^2}+\frac{r_+^4}{L^4}\right)\right]
+{\cal O}\big(\hat{\gamma}^2\big)\,, \ee where $r_+=r_H$ is the
black hole horizon. The corresponding Hawking temperature $T_H$ is 
given by 
\be \label{higher5}
T=\frac{1}{\beta}=
\frac{1}{4\pi}\,\left[e^{2\phi(r)}\right]^\prime\arrowvert_{r=r_+}
=\frac{1}{\pi\,L^2}\left[r_++\frac{k\,L^2}{2r_+}
-\frac{\hat{\gamma}}{L^2}\left(r_++\frac{2 kL^2}{r_+}
+\frac{k^2L^4}{r_+^3}\right)\right]+{\cal
O}\big(\hat{\gamma}^2\big)\,. \ee 
The black hole solutions with $k=0$ are qualitatively similar to those 
of Einstein's theory, so only the $k=\pm 1$ cases are of interest. 
We plot inverse Hawking temperature versus the horizon radius in
Fig.~(\ref{figure10}) with $k=+1$ at some fixed values of 
$\hat{\gamma}/L^2$, and in Fig.~(\ref{figure11}) but with $k=-1$. 
A Hawking-Page phase transition can occur only when the 
AdS-Schwarzschild solutions possess 
both phases: $F>0$ (thermal AdS) and $F<0$ (black hole). 
For $k=+1$, there can occur such a transition, 
driven by finite size effects, in particular, when 
$\hat{\gamma}/L^2<(0.15)/(2.511)^2$. This implies the existence of 
$F>0$ region within a certain range of $r_H$, see Fig.~(\ref{figure12}). 

\begin{figure}
%\centerline{\epsfig{file=gbblack12b.eps,width=10cm}}
\includegraphics[width=10cm]{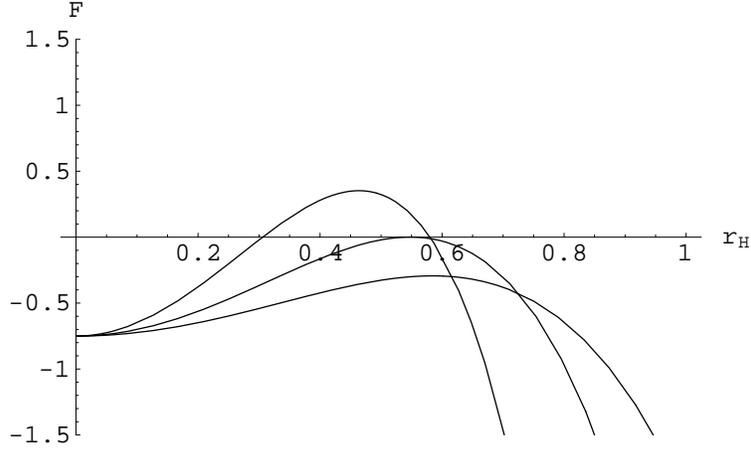}
\caption{\it Black hole in $(\mbox{Riemann})^2$ gravity:
free energy vs  horizon radius for $k=-1$ in $d=4$. 
The three curves from up to down correspond respectively to 
$\hat{\gamma}/L^2=(0.15)/(0.4)^2,~(0.15)/(0.5475)^2$ and 
$(0.15)/(0.7)^2 $.}\label{figure13}
\end{figure}

From Fig.~(\ref{figure11}) we see that the thermodynamic properties 
of topological black holes $(k=-1)$ are qualitatively similar to 
those of the five dimensional 
Gauss-Bonnet black holes with a spherical event horizon $(k=1)$. This 
behavior, however, disappears for a small $\hat{\gamma}/L^2$, for example, 
when $\hat{\gamma}/L^2<(0.15)/(0.55)^2$, which actually corresponds to the 
$F<0$ region in Fig.~(\ref{figure13}).    
     
In the high temperature limit $T>>1/L$ (i.e. $r_+ >> L$), the horizon radius 
may be expressed as 
\be \label{rintermsT}
r_+=\frac{\pi L^2\,T}{2(1-\delta)}\left[1+\sqrt{1-\frac{2(1-\delta)
(1-4\delta)k}{\pi^2\,L^2 T^2}+k^2\,\delta\,{\cal O}\left(\frac{1}{L^4T^4}
\right)}\right]\,,
\ee 
where $\delta=\hat{\gamma}/L^2\equiv 1/(8N)$. For $\delta=0$ (or 
$N\to \infty$), one has the known results of Einstein's theory: 
$T_{min}=\sqrt{2k}/(\pi L)$ and $T_c=3\sqrt{k}/(2\pi L)$. The 
$(\mbox{Riemann})^2$ correction, therefore, makes $T_{min}$ smaller 
and $T_c$ larger.    

We follow the Euclidean prescription~\cite{Hawking77a,Hawking83a,Witten98a} 
for regularizing the action and identify the Euclidean action with the 
free energy times $1/T_H$. For $(d+1)=5$,
we find the following expression for free energy~\footnote{This
result differs in sign for the last term from that
of~\cite{Nojiri01b}. Perturbative black hole solution with $R^2$ terms and
its thermodynamic behavior was further discussed in~\cite{Nojiri01c} but 
we deserve some differences in the results so far reported.}
\bea \label{higher3} 
F&=&
\frac{V_3}{\kappa_5}\,\left[\left(1-\frac{2\hat{\gamma}}{L^2}\right)
\left(\tilde{M}-\frac{2\,r^4}{L^2}\right)
-\frac{6\hat{\gamma}\,\tilde{M}^2}{r_H^4}
\right]+{\cal O}\left(\gamma^2\right)\nn \\
&=&\frac{V_3 r_+^2}{\kappa_5}\left[k-\frac{r_+^2}{L^2}
-\frac{12\hat{\gamma}k}{L^2}-\frac{3\hat{\gamma}r_+^2}{L^4}-\frac{5\hat{\gamma}
k^2}{r_+^2}\right]+{\cal O}\big(\hat{\gamma}^2\big)
\eea where we have substituted Eq.~(\ref{higher4}) in the second 
step above. In
Figs.~(\ref{figure12}) and~(\ref{figure13}), free energy of the black 
hole vs horizon radius is plotted respectively for $k=+1$ and $k=-1$. 
For $k=+1$, existence of a thermal AdS phase ($F>0$) 
requires $\hat{\gamma}/L^2<(0.15)/(2.511)^2(=1/42)$, which 
implies that $N>5$. There is no phase transition even for 
a closed geometry ($k=+1$) when $\hat{\gamma}/L^2>(0.15)/(2.511)^2$, 
and hence $N\leq 5$, because in this limit free energy is always negative. 
The existence of a minimum $N$ for the possible Hawing-Page 
phase transition is not in any contradiction. There 
is no restriction for taking a large $N$, rather this limit more and more 
ensures existence of the both phases: $F>0$ and $F<0$. As $N$ 
is increased, $\hat{\gamma}/L^2=1/(8N)$ gets decreased, and hence an 
AdS phase exists for small $r_H$, while a stable black hole phase always 
exists for large $r_H$. For $k=-1$, $F>0$ requires that $N$ be bounded 
from up, i.e. $\hat{\gamma}/L^2>(0.15)/(0.5475)^2(=1/2)$, which does not 
give a physical bound for $N$. 
%%%%%%%%%%%%%%%%%%%%%%%%%%%%%%%%%%%%%%%%%
%%%%%%%%%%%%%%%%%%%%%%%%%%%%%%%%%%%%%%%%
%%%%%%%%%%%%%%%%%%%%%%%%%%%%%%%%%%%%%%%%%%%%%% 

In order to exhibit the finite coupling effects, we may express the 
free energy density, read from Eq.~(\ref{higher3}), in terms of 
$T$ and $N$. We can define the regularized action as 
$\Delta I=V_3\beta \Delta F$, where $V_3$ is the coordinate volume of 
$\tau~(\equiv -it)$ and $d\Omega_3(k)$, and $\Delta F$ is the free energy 
density. Then, since $\hat{\gamma}/L^2=1/(8N)$ from~(\ref{sugrarel.}), 
using~(\ref{rintermsT}) and~(\ref{higher3}) we get   
\be \label{FEDwithg}
\Delta F=-\frac{\pi^2\,T^4}{4}
\left[\left(N^2+\frac{7N}{8}\right)-\left(3N^2-\frac{3N}{4}\right)
\frac{k\,\beta^2}{\pi^2 L^2}+\left(\frac{3N^2}{2}-\frac{23N}{16}\right)
\frac{k^2\,\beta^4}{\pi^4 L^4}+{\cal O}\left(\frac{\beta}{L}\right)^6
\right]\,.
\ee 
It is not a priori clear what is the finite temperature field theory 
to which this corresponds. In view of the conjectured duality between 
a string theory on 
$AdS_5\times S^5/Z_2$ background and ${\cal N}=2$ supersymmetric 
$Sp(N)$ gauge theory, this result may be relevant to the 
$N\to \infty$ (large 't Hooft coupling $g^2 N$) limit of 
the ${\cal N}=2$ Sp(N) gauge theory at finite temperature, 
which is not known. One may therefore wish to 
compare~(\ref{FEDwithg}) with those of ${\cal N}=2$ ~Sp(N) gauge 
theory (which has $4(N(2N+1)+(2N^2+7N-1))$ boson-fermion pairs) at weak 
coupling (or free field limit $g^2 N\to 0$): 
\be \label{weakcoup}
\Delta F=-\frac{\pi^2\,T^4}{3}\left(N^2+2N-\frac{1}{4}\right)\,.
\ee 
To the leading order, a difference of factor $3/4$ is seen between the 
weak and strong coupling limits. A possible explanation for the 
origin of this difference on $AdS_5\times S^5$ background, 
including the $(\mbox{Weyl})^4$ terms, was first 
given in~\cite{Gubser98a}, 
and further comments on this difference were made in the 
review~\cite{Gubser00a} (see also references therein). It is expected that 
a similar relation would apply to the string theory on 
$AdS_5\times S^5/Z_2$ background. Note that even on the flat 
space $k=0$, two results do not seem to match each other, the role is 
here being played by $(\mbox{Riemann})^2$ term. 
%%%%%%%%%%%%%%%%%%%%%%%%%%%%%%%%%%%
%One does not 
%expect that deformations at finite temperature on $AdS_5\times S^5/Z_2$ 
%background lead only to marginal deformation (which does not break 
%conformal invariance to leading order) in the gauge theory. 
%%%%%%%%%%%%%%%%%%%%%%%%%%%%%%%%%%%%%%%%%%%%%%%%%%%%% 

By considering a 5d AdS Schwarzschild background, but 
with a trivial $(\mbox{Riemann})^2$ bulk term, and using AdS/CFT 
correspondence, we can calculate the free energy density, which is 
relevant to the $N\to \infty$ limit of the ${\cal N}=4$ SYM 
gauge theory at finite temperature~\cite{Witten98a}. This 
reads   
\be \label{FEDwithg0}
\Delta F=-\frac{\pi^2 N^2 T^4}{8}\left[1-\frac{3\,k\,\beta^2}{\pi^2 L^2} 
+\frac{3}{2\pi^4}\,\frac{k^2\beta^4}{L^4}+{\cal O}
\left(\frac{\beta}{L}\right)^6\right]\,.
\ee 
While, the high temperature limit of the free energy density for four 
dimensional ${\cal N}=4$ supersymmetric SU(N) theory at weak 
coupling is given 
by~\cite{Burgess99a}  
\be \label{FEDwithT}
\Delta F=-\frac{\pi^2 N^2 T^4}{6}\left[1-\frac{3\,k\,\beta^2}{2\pi^2 L^2} 
+{\cal O}\left(\frac{\beta}{L}\right)^6\right]\,.
\ee 
There is already a mismatch in the sub-leading terms 
between~(\ref{FEDwithg0}) and~(\ref{FEDwithT}). 
It would be not surprising if similar situation arises for ${\cal N}=2$ 
Sp(N) gauge theory. There is also a temperature independent term 
in~(\ref{FEDwithg0}), which is interesting to study further. A difference 
of factor $1/2$ in the leading term for free energy density 
between~(\ref{FEDwithg}) and~(\ref{FEDwithg0}) simply 
arises from the dictionary of AdS/CFT: in SU(N) SYM theory 
one uses $G_5L^5=8\pi^3 g_{YM}^2{\alpha'}^4=\pi L^8/(2N^2)$ which differs 
by $1/2$ from~(\ref{sugrarel.}) due to a volume factor 
Vol($S^5/Z_2$) versus Vol($S^5$).
 
To evaluate thermodynamic quantities of the black holes, including 
entropy, we go back to the results~(\ref{higher5}) 
and~(\ref{higher3}). The entropy of the black hole can be determined 
by using the relation 
\be {\cal S}=-\frac{dF}{dT_H}=-\frac{dF}{dr_+}\,
\frac{dr_+}{dT_H}\,, \ee where the derivative terms $dF/dr_+$ and
$dT_H/dr_+$ take the following forms \bea \label{higher7}
\frac{dF}{dr_+}&=&-\frac{4V r_+^3}{\kappa_5\,L^2}
\left[1-\frac{k\,L^2}{2r_+^2}+\frac{3\hat{\gamma}}{L^2}
\left(1+\frac{2k\,L^2}{r_+^2}\right)\right]+{\cal
O}\big(\hat{\gamma}^2\big)\,
,\nn \\
\frac{dT_H}{d
r_+}&=&\frac{1}{\pi\,L^2}\left[1-\frac{k\,L^2}{2r_+^2}
-\frac{\hat{\gamma}}{L^2}\left(1-\frac{2kL^2}{r_+^2}
-\frac{3k^2L^4}{r_+^4}\right) \right]\,+{\cal
O}\big(\hat{\gamma}^2\big)\,. \eea Hence we obtain (since
$\hat{\gamma}=2\gamma\,\kappa_5$) 
\be \label{higher8} 
{\cal S}=\frac{4\pi V_3
r_H^3}{\kappa_5}\left[1+\frac{8\gamma\,\kappa_5}{L^2}
\left(1+\frac{k\,L^2}{r_+^2}-\frac{3k^2 L^4}{4\, r_+^4}\right)
\left(1-\frac{k\,L^2}{2\,r_+^2}\right)^{-1}\right]+{\cal
O}(\hat{\gamma}^2)\,. \ee 
For a large size black hole $2r_+^2>>
kL^2$, one may approximate the entropy
\be \label{entropyapprox}
{\cal
S}\simeq\frac{V_3 r_+^3}{4\,G_5}\left[1
+\frac{8\gamma\,\kappa_5}{L^2}
\left(1+\frac{3k\,L^2}{2\,r_+^2}\right)\right]+{\cal
O}(\gamma^2)\,.
\ee Thus entropy of the black hole with
$(\mbox{Riemann})^2$ term does not satisfy the area formula ${\cal
S}=A/4G$. Rather, just as the Einstein term in the action is
corrected by $(\mbox{Riemann})^2$ term, the Einstein contribution
to black hole entropy (${\cal S}=A/4G$) receives
$(\mbox{Riemann})^2$ corrections. This has been expected in the
literature (see Ref.~\cite{Visser93a} for review), but here we further
show that entropy~(\ref{entropyapprox}) agrees with Wald's formula for
entropy~\cite{Wald93a} for $r_+^2>>L^2$. Moreover, we may implement 
these results to reproduce boundary data on the horizon of the AdS 
geometry.

We can also find the black hole energy, defined as $E~(\equiv
T\,{\cal S}+F)$, \be \label{BHenergy}
E=\frac{3\,V_3}{\kappa_5}\left[\tilde{M}_0+2\gamma\kappa_5
\left(\frac{3\,r_+^4}{L^4}+\frac{5 k\,r_+^2}{2 L^2} -k^2-\frac{k^3
L^2}{2 r_+^2}\right)
{\left(1-\frac{kL^2}{2\,r_+^2}\right)}^{-1}\right]+{\cal
O}\big(\gamma^2\big) \,. \ee Here
$\tilde{M}_0=r_+^2\left(k+r_+^2/L^2\right)$ is the value of
$\tilde{M}_+$ when $\gamma=0$. For $k=1$, one apparently sees a
singularity at $2r_+^2=L^2$, but there is no singularity
in the formulas for energy Eq.~(\ref{BHenergy}) and entropy
Eq.~(\ref{higher8}). One easily checks that at $r_+^2= L^2/2$, for
$k=1$, the first round bracket in~(\ref{BHenergy}) also vanishes,
thus the above formulas are applicable to all three possible 
values: $k=0,\,\pm 1$.

It has been known that entropy of the black hole can be expressed
as a local geometric (curvature) density integrated over a
space-like cross section of the horizon. Notably, an entropy
formula valid to any effective gravitational action including higher
curvature interactions was first proposed in~\cite{Wald93a}, and was 
nicely generalized in Refs.~\cite{Jacobson95a,Myers98a}. One can infer
from~\cite{Myers98a} that the black hole entropy for the
action~(\ref{action1}) takes in five dimensions the following form
\be\label{higher9} \tilde{{\cal
S}}=\frac{4\pi}{\kappa_5}\int_{horizon} d^3x
\sqrt{h}\,\bigg\{1+2\alpha\kappa_5\,R+\beta\kappa_5 \left(R-h^{ij}
R_{ij}\right) +2\gamma\,\kappa_5\Big(R-2h^{ij}R_{ij}+h^{ij} h^{kl}
R_{ikjl}\Big)\bigg\}\,, 
\ee 
where $h$ is the induced metric on the
horizon. For $d+1=5$, since the curvatures are defined in the
following form 
\be R=-\frac{20}{L^2}\,,\quad
h^{ij}R_{ij}=-\frac{12}{L^2}\,, \quad h^{ij}
h^{kl}R_{ikjl}=\frac{6k}{r_+^2} +{\cal O}(\hat{\gamma})\,, 
\ee
from the Eq.~(\ref{higher9}), when $\alpha=\beta=0$, we read
\be\label{higher10} 
\tilde{{\cal S}}=\frac{V_3\,r_+^3}{4G_{5}}
\left[1+\frac{8\gamma\,\kappa_5}{L^2}\left(1+\frac{3k\,L^2}
{2\,r_+^2}\right)\right]+{\cal O}(\gamma^2)\,. 
\ee 
Hence the two
expressions for entropies, i.e. Eq~(\ref{higher10}) and
Eq.~(\ref{entropyapprox}), are identical. Our results are suggestive, and
clearly contradict with the observations made in
Ref.~\cite{Nojiri01b} in this regard.

From the $(d+1)$ dimensional analogue of the
formula~(\ref{higher9}), we may calculate entropy of the Gauss-Bonnet
black hole, using the relation $\alpha=-\beta/4=\gamma$, \be
\tilde{{\cal S}}=\frac{4\pi}{\kappa_{d+1}} \oint
d^{d-1}x\,\sqrt{h}\left[1+2\alpha\kappa_{d+1}{\cal R}(h)\right]\,.
\ee At the horizon, one sets $e^{2\phi(r_+)}=0$, and reads the value
of ${\cal R}(h)$ from~(\ref{Rscalar}). Hence \be \label{GBENTROPY}
{\cal
S}=\frac{V_{d-1}r_+^{d-1}}{4G_{d+1}}\left[1+\frac{(d-1)}{(d-3)}\,
\frac{2\hat{\alpha}\,k}{r_+^2}\right]\,, \ee which coincides with
Eq.~(\ref{entropyGB}). Myers and Simon~\cite{Myers88a} have
derived the result~(\ref{GBENTROPY}) for entropy of the black hole
in asymptotically flat backgrounds ($L^2\to \infty$ or
$\Lambda=0$) with $k=1$, but here we see that this holds for arbitrary
$\Lambda$. This mimics that the geometry on the boundary or the cosmological
term in the bulk is not dynamical. This is plausible and possibly
gives some insights of the holography.

\subsection{Quantities with non-trivial $\alpha\,,\beta\,,\gamma$}

Furthermore, one could find a perturbative solution for arbitrary $\alpha$,
$\beta$ and $\gamma$ at a time, but this calculation is complicated
due to a perturbative expansion. Without loss of any generality,
one can follow a different, but equivalent, prescription, in which one
combines the results obtained for (i) $\alpha,\beta\neq 0$ and
$\gamma=0$, to the results obtained for (ii) $\alpha=\beta=0$ and
$\gamma\neq 0$. Thus, by combining the
results~(\ref{freeFforg0},\ref{energy1})
and~(\ref{higher3},\ref{higher8},\ref{BHenergy}), we obtain the
following expressions for free energy $(F)$, entropy $({\cal S})$
and energy $(E)$ of the $5d$ AdS black hole with quadratic
curvature terms
 \be \label{freeenergy}
F=\frac{V_3}{\kappa_5}\left[r_H^2\left(k-\frac{r_H^2}{L^2}\right)
\left(1-\varepsilon\right)-10\gamma\kappa_5\left(k+\frac{r_H^2}{L^2}\right)^2
\,\right]+{\cal O}\big(\gamma^2\big)\,, \ee \be\label{higher11}
{\cal S}=\frac{4\pi V_3\,r_H^3}{\kappa_5}
\left[(1-\varepsilon)+\frac{12\gamma\,\kappa_5}{L^2}\left(1+\frac{k\,L^2}
{2\,r_H^2}-\frac{k^2 L^4}{2r_H^4}\right) \left(1-\frac{k L^2}{2
r_H^2}\right)^{-1}\right] +{\cal O}(\gamma^2)\,, \ee \be
\label{totalE}
E=\frac{3\,V_3}{\kappa}\bigg[\tilde{M}_0\big(1-\varepsilon\big)
+2\gamma\kappa_5\left(\frac{5 r_H^4}{L^4}+\frac{7\,k
r_H^2}{2\,L^2} -2 k^2-\frac{k^3 L^2}{2 r_H^2}\right)
\left(1-\frac{kL^2}{2r_H^2}\right)^{-1}\bigg]+{\cal
O}\big(\gamma^2\big)\,, \ee where
$\varepsilon=4\kappa_5\left(\gamma+2\beta+10\alpha\right)/L^2$, and
$\tilde{M}_0$ is related to the mass parameter $\tilde{M}$ when
$\alpha=\beta=\gamma=0$. The consistency of the above results is reflected
from the regularities of the expressions for entropy~(\ref{higher11})
and energy~(\ref{totalE}) at $kL^2=2r_H^2$.

In the large $N$ limit, one has $L^3>>\kappa_5$ (from the first expression of
Eq.~(\ref{sugrarel.})), and in AdS/CFT, the coefficients of the higher
curvature terms in the bulk, in general, satisfy the limit
$1> \varepsilon > 0$. In Ref.~\cite{Blau99a} the subleading
contribution to the $AdS_5/CFT_4$ trace anomaly was considered. In
terms of boundary metric $g_{ij}^{(0)}$, supergravity prediction of
${\cal O}(N)$ contribution to the trace anomaly was found to
be~\cite{Blau99a}
\be
\frac{6N}{24\times 16\pi^2}\left[R_{ijkl}^{(0)}{R^{(0)}}^{ijkl}
-\frac{13}{4}\,R_{ij}^{(0)} {R^{(0)}}^{ij}+\frac{3}{4}\,{R^{(0)}}^2\right]
\,,
\ee
from where we see that $1<\varepsilon<0$ holds in large $N$ limit. 
Since $\gamma>0$, the free energy~(\ref{freeenergy})
is always negative for $k=0$ and $k=-1$, and such black holes
are globally stable. This means that provided
$(1-\varepsilon) >0$ and $\gamma>0$, there may not occur a
Hawking-Page phase transition for AdS black hole with a Ricci flat
or hyperbolic horizon for the above theory. For the $k=1$
case, $F$ can be negative for
large black hole with $r_H^2>> L^2$ since $\gamma>0$. While, for a small
size black hole $r_H^2<L^2$, the free energy $F$ can be positive.

As a consistency check of the above formulas, we can express
the curvature squared terms in the Gauss-Bonnet form:
$\alpha=-\beta/4=\gamma$. In this case, since
$\varepsilon=12\alpha \kappa/L^2$, the free
energy~(\ref{freeenergy}) reduces to \be \label{freeenergyGB}
F=\frac{V_3}{\kappa_5}\left[r_+^2\left(k-\frac{r_+^2}{L^2}\right)
+\hat{\alpha}\left(\frac{r_+^4}{L^4}-\frac{16\,k r_+^2}{L^2}-5
k^2\right) \right]+{\cal O}\big(\alpha^2\big)\,, \ee where
$\hat{\alpha}=2\alpha\kappa_5$ and $r_+=r_H$. This agrees with
the exact expression of
free energy~(\ref{GBfreeE}), where one sets $d=4$ and
$(1+2\hat{\alpha}k/r_+^2)^{-1}\equiv (1-2\hat{\alpha}k/r_+^2)$, and
$L\equiv \ell$. We may read entropy for the Gauss-Bonnet
black hole from the formula Eq.~(\ref{higher11}), in the limit
$r_+^2>>L^2$, \be\label{higher12} {\cal S}_{GB}=\frac{4\pi
V_3\,r_+^3}{\kappa_5}
\left[1+\frac{12\alpha\kappa_5\,k}{r_+^2}\right]\,. \ee This very
nicely agrees in five dimension ($d=4$) with the
expression~(\ref{entropyGB}) or~(\ref{GBENTROPY}) (obtained using two 
different prescriptions). For $k=0$, one has
${\cal S}_{GB}=(4\pi V_3\,r_H^3)/\kappa_5\equiv A/4G$. This special
connection is the
reminiscent of the topological behavior of the GB invariant, but as we 
have already seen ${\cal S}=A/4G$ does not hold in the generic 
higher derivative theories.

\section{Holography beyond AdS/CFT}

Via holography~\cite{tHooft93a}, it has been known that
thermodynamic quantities of a boundary CFT can be determined by those
of the global AdS (supergravity) vacuum, the notion of the
celebrated AdS/CFT correspondence~\cite{Maldacena97a}.
Witten~\cite{Witten98a} further argued that such a correspondence
may exist even if we give a finite temperature 
to the bulk AdS (so that pure AdS bulk is replaced by
AdS-Schwarzschild black hole), and define a CFT on the boundary 
at finite temperature. Of course, finite temperature breaks both 
the supersymmetry and the conformal invariance. Still one can associate 
the mass (energy), temperature and entropy of the black hole with the
corresponding quantities in the boundary field theory at finite 
temperature~\cite{Witten98a}. Many checks that have been performed 
in the literature 
are either based on the Einstein's theory or limited in the large $N$ 
limit, where one neglects any higher curvature corrections. However, when 
one includes a bulk $(\mbox{Riemann})^2$ term, and extracts boundary 
data (e.g. free energy, entropy, energy) from AdS-Schwarzschild 
solution, one finds that they do not agree to the sub-leading 
order. This we will exhibit below in a simple cosmological context. 

\subsection{Relating boundary and bulk parameters}
To reproduce boundary data from bulk data
and vice versa, one may consider the coordinate transformations
given in the footnote~(\ref{co-trans}), which are solved
by~\cite{Nojiri01a} \be
\label{relate1}
{\sigma^*}^2={\sigma'}^2-e^{2\phi}(r)\,e^{2\sigma(y)}\,, \ee
where $d\eta=e^{-\sigma(y)} d\tau$, with $\eta$ being a new time
parameter, $\sigma^*=\partial_\eta\sigma$ and $\sigma'=\partial_y\sigma$.
One specifies the functions $r=r(\eta)$, $t=t(\eta)$, so that
$-\sigma^*\equiv \dot{r}/r$ defines the Hubble parameter $H$.
Eq.~(\ref{relate1}) ensures that the induced
metric on the brane takes the standard Robertson-Walker form \be
ds_d^2=-d\eta^2+r^2(\eta)\,d\Omega_{d-1}^2\,, \ee
and the radial distance $r$ measures the size of $d$-dimensional
universe from the center of the black hole~\cite{Verlinde00a}. When applying
the holography in the above context, one could consider an $d$-dimensional
brane with a constant tension in the background of an $(d+1)$-dimensional
AdS black hole. One also regards the brane as the boundary of the AdS
geometry, and further assumes that ${\sigma'}|_{y=0+}=- L^{-1}$ at the horizon
$r=r_H$, where $e^{2\phi(r)}=0$.
Then it is clear that Eq.~(\ref{relate1}) leads to $H=\pm 1/L$. Using
Eq.~(\ref{HDbrane4}), one finds the first Friedman equation in
$(d+1)$-dimensions~\cite{Nojiri01b}
\be\label{relate2}
H^2+\frac{k}{r^2}= \frac{\tilde{M}}{r^d}\equiv
\frac{\kappa_d}{(d-1)(d-2)}\, \rho\,, \ee where $\kappa_d=16\pi
G_{(d)}$, $\rho=\tilde{E}/V$ is the matter energy density on the boundary,
$V=V_{d-1} r^{d-1}$, and \be\label{relate3}
\tilde{E}=\frac{(d-1)(d-2)\tilde{M}\,V_{d-1}}{\kappa_d\,r}\,.
\ee 
The energy $\tilde{E}$ coincides with the
gravitational energy $E$ (Eq.~(\ref{energy1})), up to a
conformal factor. Differentiation of Eq.~(\ref{relate2}) gives the second
Friedman equation
\be
\dot{H}-\frac{k}{r^2}=-\frac{\kappa_d}{2(d-2)}\,\big(\rho+p\big)\,,
\quad \mbox{where}\quad p=\frac{(d-2)\,\tilde{M}}{\kappa_d\,r^d}
\ee
is the matter pressure on the brane. Since $-\rho+(d-1)\,p=0$, the
induced CFT matter is radiation-like $T^\mu_\mu=0$. For a non-zero
$\gamma$ (but with $\alpha=\beta=0$), however, the FRW equation reads
\be \label{FRW1}
H^2+\frac{k}{r^2}=\frac{\kappa_4}{6}\,\rho\,,
\ee
where $\rho=\tilde{E}/V$, $V=V_3\,r^3$ and
\be \label{energy2}
\tilde{E}=\frac{6V_3}{\kappa_4\,r}\left[\tilde{M}
-\frac{\hat{\gamma}\tilde{M}^2}{r^4}\right]\,.
\ee
Differentiation of Eq.~(\ref{FRW1}) gives the
second FRW equation
\be \label{FRW1withg}
\dot{H}-\frac{k}{r^2}=-\frac{\kappa_4}{4}\left(\frac{\tilde{E}}{V}+p\right)
\,,
\ee
with
\be \label{pressure2}
p=\frac{2}{\kappa_4\,r^4}\left[\tilde{M}
-\frac{5\hat{\gamma}\tilde{M}^2}{r^4}\right]\,.
\ee
From Eqs.~(\ref{energy2}) and~(\ref{pressure2}), one has 
\be \label{bigL}
-\frac{\tilde{E}}{V}+3p=
-\frac{24\hat{\gamma}\,\tilde{M}^2}{\kappa_4\,r^8}
+{\cal O}\left(\hat{\gamma}^2\right)\,.
\ee 
This simply means that matter fields on the boundary (brane) does not satisfy 
a radiation condition ($T_\mu^\mu=0$) for $\gamma\neq 0$. This is not an 
unexpected behavior with a bulk $(\mbox{Riemann})^2$ term. To proceed 
further, it is desirable to define the relation between $\kappa_4$ and 
$\kappa_{5}$.
They can be related by (see~\cite{Gubser99a} for $\varepsilon=0$ case)
\be \label{twokappas}
\kappa_4=\frac{2}{L}\,\frac{\kappa_{5}}{1+\lambda \varepsilon}\,.
\ee
If $\alpha=\beta=0$ and $\gamma\neq 0$, one has
$\varepsilon=4\kappa_5\gamma/L^2$. The magnitude of
$\lambda$ can be fixed from the propagator analysis. If one regards
the brane as the boundary of the AdS geometry, one has
$\lambda=1$. We simply note, however, that for a gravity localized
Randall-Sundrum type $\delta$-function brane, one actually finds
$\lambda=1/3$ (see Ref.~\cite{IPN01b,IPN01e} for the Gauss-Bonnet case).
Assume that brane is the horizon of AdS geometry, then it is suggestive
to consider a moment in the brane's cosmological evolution at which
the brane crosses the black hole horizon $r=r_{brane}=r_H$, so that
$\tilde{M}(r_{brane})=\tilde{M}(r_H)$. After implementing the
relation~(\ref{twokappas}), i.e.
\be \label{kappa4&5}
\kappa_4=\frac{2}{L}\,\frac{\kappa_5}{1+2\hat{\gamma}/L^2}\equiv
\frac{2\,\kappa_5}{L}\,\left(1-\frac{2\hat{\gamma}}{L^2}\right)\,,
\ee
from Eq.~(\ref{energy2}) we find
\be \label{braneE}
\tilde{E}=\frac{3V_3}{\kappa_5}\,\frac{L}{r}
\left[\tilde{M}_0+\frac{4\gamma\kappa_5\,r_+^2}{L^2}
\left(k+\frac{r_+^2}{L^2}\right)\right]
+{\cal O}\big(\gamma^2\big)\,.
\ee
One may assume that energy $\tilde{E}$ is rescaled by
\be \label{twoEs}
\tilde{E}=\frac{L}{r}\,E_{AdS}\,.
\ee
The origin of the factor $L/r$ is entirely ``holographic'' in spirit.
It has been argued in
Ref.~\cite{Padilla01a} that for non-critical
branes (i.e. $k\neq 0$), the AdS length scale $L$ may be replaced by
$1/\T$, where $\T$ is the brane tension. At any rate, we
see that the two expressions for energy, Eq.(\ref{BHenergy}) and
$E_{AdS}$ (which one reads from Eqs.~(\ref{braneE}) and~(\ref{twoEs}))
do not match to the subleading order, but for $\gamma=0$ they do coincide. 

Finally, we comment upon the size and location of the brane for a flat
and static brane. Instead of $\sigma^\p=-1/L$, let us assume that
$\sigma^\p=-1/\ell$ holds. If the brane (hypersurface) is flat ($k=0$),
the horizons defined by $e^{2\phi(r)}=0$ (Eq.~(\ref{higher1a})) read
\be
r_+^4=\mu\,l^2+\frac{\mu\hat{\gamma}}{3}+{\cal O}\big(\hat{\gamma}^2\big)\,,
\quad r_-^4=\mu\hat{\gamma}
\left(1-\frac{\hat{\gamma}}{l^2}-\frac{16}{3}\,\frac{\hat{\gamma}^2}{l^4}
\right)+{\cal O}\big(\hat{\gamma}^3\big)\,.
\ee
If the brane is also static (i.e. $H=0$), the Friedmann
equation~(\ref{FRW1withg}) would rise to give two horizons.
However, the one which reduces to that of the Einstein gravity 
for $\gamma=0$ and would be of more physical interest is
\be
r_{brane}^4=\mu\hat{\gamma}\left(1-\frac{2\hat{\gamma}}{l^2}
+\frac{2}{3}\,\frac{\hat{\gamma}^2}{l^4}\right)
+{\cal O}\big(\hat{\gamma}^3\big)\,.
\ee
Since $\gamma>0$, there is a critical point at
\be
1=\frac{6\hat{\gamma}}{l^2}\equiv \frac{12\,\gamma\kappa_5}{l^2}\,,
\ee
where the brane coincides with the black hole inner horizon
$r_{brane}^4=r_-^4$. The brane lies outside the (black hole) inner
horizon if $l^2> 6\hat{\gamma}$, and it lies inside the black hole horizon
if $l^2< 6\hat{\gamma}$.

\subsection{Entropy bounds in holography}
For completeness and comparison, we list some of the interesting
proposals for entropy bounds in holography~\cite{Rey99a,Youm01a}.
Consider the Cardy formula~\cite{Cardy86a} of two-dimensional
conformal field theory \be \label{cardy} {\cal
S}_H=2\pi\sqrt{\frac{c}{6}\,\left(L_0-\frac{c\,k}{24}\right)} \,,
\ee where the CFT generator (eigenvalue) $L_0$ represents the
product $E\,r$ of the energy and radius, $k$ is the
intrinsic curvature of the CFT boundary, $c/24$ is the shift in
eigenvalues caused by the Casimir effect.
In~\cite{Verlinde00a,Savonije01a}, it was shown that the
formula~(\ref{cardy}) can be generalized to arbitrary $d$-dimensions,
which then corresponds to the FRW type brane equations of $d$
dimensions. One identifies there~\cite{Savonije01a,Youm01a} 
\be
\label{Verlinde} 
L_0\equiv \frac{E\, r}{d-1}\,,\quad
\frac{c}{6}\equiv \frac{4(d-2) V_3\,r^2}{\kappa_d}\,, \quad {\cal
S}_H\equiv \frac{4\pi(d-2)\,H V}{\kappa_d}\,, 
\ee where ${\cal
S}_H$ is the Hubble entropy~\cite{Verlinde00a}. The Cardy formula
puts a bound ${\cal S}_H <{\cal S}_B$, where \be {\cal
S}_B=\frac{2\pi\,r}{d-1}\,{E} \ee is the Bekenstein entropy.
The ``holographic bound'' proposed by 't Hooft and Susskind reads
${\cal S}\leq {\cal S}_{BH}$, where \be {\cal S}_{BH}=\frac{4\pi\,
(d-2)}{\kappa_d\,r}\,V\,. \ee This mimics that entropy is smaller
than Bekenstein-Hawking entropy of the largest black hole that
fits in the given volume. Another entropy bound is the ``Hubble
bound'' proposed in~\cite{Rey99a}, which reads ${\cal S}\leq S_H$.
This heralds that entropy is bounded by total entropy of Hubble
size black hole.

For our purpose, we find entropy formula defined
in~(\ref{Verlinde}) interesting. Though initially derived for
two-dimensional CFT, this formula may be applied to
$CFT_4/AdS_{5}$~\cite{Verlinde00a}, then one regards ${\cal S}_H$ 
as the holographic entropy. One may worry about when applying this
formula to a theory with $(\mbox{Riemann})^2$ term.
We therefore follow here a heuristic approach to estimate a difference
between ${\cal S}_H$ and ${\cal S}_{BH}$. Note that, for $\gamma\neq 0$, a
non-conformal behavior in entropy is seen also from the scaling relation
between
$\kappa_4$ and $\kappa_5$, the former one involves a non-trivial
$\gamma$ via~(\ref{kappa4&5}). Using the value $H_0=1/L$ and
Eq.~(\ref{kappa4&5}), the Hubble entropy ${\cal S}_H$
in~(\ref{Verlinde}) takes the form, for $d=4$, 
\be \label{HubbleS}
{\cal S}_{H}=\frac{4\pi V_3\,r_H^3}{\kappa_5}
\left(1+\frac{4\gamma\,\kappa_5}{L^2}\right)
\equiv\frac{V}{\pi L^3}\left(N^2+\frac{N}{4}\right)\,. \ee 
Of course, in the large $N$ limit the correction term is negligible. 
Nevertheless, we may 
calculate the difference in Bekenstein-Hawking entropy ${\cal S}_{BH}$
(Eq.~(\ref{higher8})) and Hubble entropy (Eq.~(\ref{HubbleS})),
which reads \bea {\cal S}_H-{\cal S}_{BH} &=&\frac{4\pi\,V_3
r_H^3}{\kappa_5} \left[-\frac{4\kappa_5\,\gamma}{L^2}\,
\left(1+\frac{5}{2}\,\frac{kL^2}{r_H^2}-\frac{3}{2}\,\frac{k^2
L^4}{r_H^2}
\right)\left(1-\frac{k\,L^2}{2\,r_H^2}\right)^{-1}\right] \nn \\
&=&- 48\pi V_3\gamma\,r_H \left(k+\frac{r_H^2}{3 L^2}\right)
\equiv -\frac{3N\,V_3\,r_H}{4\pi\,L}\left(k+\frac{r_H^2}{3\,L^2}\right)\,.
\eea 
Note that the difference in entropies can arise only in the order of
$N$, though each entropy is proportional to $N^2$. Moreover, a 
negative sign in the difference implies
that ${\cal S}_H\leq {\cal S}_{BH}$. An obvious fact that
${\cal S}_{H}$ does not coincide with ${\cal S}_{BH}$ is not surprising
with $\gamma\neq 0$, because in higher derivative gravity entropy is 
not directly related to the horizon. The entropy formulas for the boundary 
field theory and the FRW equations coincide only for a radiation
dominated (brane) universe~\cite{Verlinde00a}. Of course, 
for $\gamma=0$, one has ${\cal S}_H={\cal S}_{BH}$. For $k=-1$, 
one has ${\cal S}_H >{\cal S}_{BH}$ if $r_H^2>3 L^2$, 
but one has $S_H<{\cal S}_{BH}$ for $L^2 < r_H^2 < 3L^2 $. In the 
$k=0$ and $k=1$ cases, one finds ${\cal S}_H\leq {\cal S}_{BH}$.
Thus for a flat and spherical AdS black holes, we propose a new entropy
bound ${\cal S}_H\leq S_{BH}$. These results could be useful to 
understand possible entropy bounds for the non-trivial $(\mbox{Riemann})^2$ 
interaction. 

\section{Conclusions}
%%%%%%%%%%%%%%%%%%%%%%%%%%%%%%%%%%%%%%%%%%%%%%%%%%%%%%%%%%%%%%%%%%%%%%%%%

A possibility is that holography beyond the AdS/CFT persists in
true quantum gravity and unified field theory, which may require 
inclusion of the higher order curvature derivative terms. With this
motivation, we studied in details the thermodynamic properties
of anti-de Sitter black holes for the Einstein-Gauss-Bonnet theory, and
the Einstein term in the action corrected by the generic $R^2$ terms,
and discussed their thermodynamic behavior including the thermal
phase transitions in AdS space. We obtained in useful form the expression
of black hole free energy for the theories with Guass-Bonnet 
and $(\mbox{Riemann})^2$ interaction terms, and calculated entropy and 
energy. In the Einstein's theory, only
the $k=1$ case is physical to explain the Hawking-Page phase
transition. This is suggestive, because the AdS solution with $k=1$ can be
embedded in ten-dimensional IIB supergravity such that the supergravity
background is of the form $AdS_5\times S^5$, and the boundaries of the
bulk manifold will have the same intrinsic geometry as the background of the
dual theory at finite temperature. We also noted that, unlike to the 
Einstein theory, in the EGB theory there may occur a Hawking-Page phase 
transition also for an event horizon of negative curvature ($k=-1$) 
hypersurface. The free energy of such a topological black hole starts 
from a negative value, reaches a positive maximum at some $r=r_+$, 
and then again goes to negative infinity as $r_+\to \infty$. Thus, 
a hyperbolic AdS black hole, though can have a thermal AdS phase within 
a small range of $r_H$, globally prefers black hole phase for large $r_H$. 

It is shown that the contribution from the squares of Ricci
scalar and Ricci tensor can be absorbed into free energy, entropy,
and energy via a redefinition of the five dimensional
gravitational constant and the radius of curvature of AdS space.
By introducing the Gibbon-Hawking surface term and a boundary
action corresponding to the vacuum energy on the brane, we
recovered the RS type fine tunings as natural consequences of the
variational principle. An interesting observation is that the 
thermodynamic properties of $k=-1$ AdS black holes, for $d+1=5$, under 
a critical value of $\hat{\gamma}/L^2$ in $(\mbox{Riemann})^2$
gravity are qualitatively similar to those of Gauss Bonnet black
hole with a spherical event horizon ($k=1$). 
For a spherical horizon geometry, the $(\mbox{Riemann})^2$ term 
(i.e. a finite coupling but $N>5$) may induce a Hawking-Page phase 
transition, though a large $N$ is much preferred. We discussed the 
corrections to the free energy density originated from the 
higher order curvature corrections to the supergravity action, which 
change only the $N$ dependence coefficients by sub-leading 
terms. Nonetheless, such corrections are necessary to look the 
behavior of AdS/CFT next to leading order, which we otherwise compare 
only to the CFTs leading terms, e.g., in the free energy and entropy. 
Moreover, in $(\mbox{Riemann})^2$-gravity, the Bekenstein-Hawking 
entropy agrees with the value obtained using
similar formula when the black hole horizon $r_H$ is much larger
than the AdS length scale $L$. The Bekenstein-Hawking entropy in
the Gauss-Bonnet theory coincides with the value directly
evaluated using Wald's formula. These are interesting and quite
pleasing results. We also obtained formulas for free energy,
entropy and energy of the AdS black holes with generic $R^2$ terms, 
which are applicable to all three geometries: $k=0,\,\pm1$. 

We have also established some relations between the boundary field theory
parameters defined on the brane and the bulk parameters associated with the
Schwarzschild AdS black hole in five dimensions. We found the FRW type 
equations on the brane, which exhibit that matter on the AdS boundary 
does not satisfy the radiation condition ($T_\mu^\mu=0$) for a non-zero 
$\gamma$. Using a heuristic but viable approach, we calculated 
a difference between the holographic entropy and the 
Bekenstein-Hawking entropy, which generally implies 
${\cal S}_{H}\leq {\cal S}_{BH}$. The essential new ingredient 
in our analysis is the role of the higher curvature terms in 
explaining the essential features of the black hole thermodynamics, 
including thermal phase transitions, finite coupling effects, and 
holography.

\section*{Acknowledgements}

This work was supported in part by the BK21 
program of Ministry of Education. One of us (IPN) also acknowledges 
partial support from the Seoam Foundation, Korea, and would like to 
thank ASICTP for hospitality where part of the work is done.  
Helpful discussions with M. Blau, Edi Gava and K. S. Narain are 
gratefully acknowledged.

%\newpage
 
\end{document}